\documentclass[a4paper,10pt]{article}
\usepackage{graphicx} % Required for inserting images
\usepackage{amsmath,amsfonts,amssymb,mathtools, hyperref, verbatim, multirow, booktabs,subcaption, float}
\usepackage[table]{xcolor}
\usepackage{relsize}

\usepackage[backend=biber, style=phys, biblabel=brakets, url=true]{biblatex}
\usepackage[skins,theorems]{tcolorbox}
\tcbset{highlight math style={enhanced,
colframe=orange,colback=white,arc=0pt,boxrule=1pt}}
\usepackage{mathrsfs}
\usepackage{tikz}
\usetikzlibrary{quantikz2}

\usepackage{makecell}

\usepackage{csquotes}

\usepackage{array}
\newcolumntype{?}{!{\vrule width 1.1pt}}

\newcommand{\Z}{\mathtt{Z}}  

\newcommand{\HH}{\mathtt{H}}

\newcommand{\RX}{\mathtt{R_X}}
\newcommand{\RY}{\mathtt{R_Y}}
\newcommand{\RZ}{\mathtt{R_Z}}

\newcommand{\R}{\mathbb{R}}

\newcommand{\N}{\mathbb{N}}

\title{Benchmarking data encoding methods in Quantum Machine Learning}
\author{Orlane Zang, Grégoire Barrué and Tony Quertier}
\date{}

\begin{document}
\maketitle

\begin{abstract}
Data encoding plays a fundamental and distinctive role in Quantum Machine Learning (QML). While classical approaches process data directly as vectors, QML may require transforming classical data into quantum states through encoding circuits, known as quantum feature maps or quantum embeddings. This step leverages the inherently high-dimensional and non-linear nature of Hilbert space, enabling more efficient data separation in complex feature spaces that may be inaccessible to classical methods. This encoding part significantly affects the performance of the QML model, so it is important to choose the right encoding method for the dataset to be encoded. However, this choice is generally arbitrary, since there is no \enquote{universal} rule for knowing which encoding to choose based on a specific set of data. There are currently a variety of encoding methods using different quantum logic gates. We studied the most commonly used types of encoding methods and benchmarked them using different datasets.    
\end{abstract}

\section*{Introduction}

Quantum Machine Learning (QML) is a research area that focuses on the development of Machine Learning (ML) algorithms that can be executed by a quantum computer. Taking advantage of quantum phenomena such as quantum superposition and quantum entanglement, the incorporation of quantum computing in ML aims to leverage the power of the quantum computer to improve existing classical ML algorithms \cite{Alvarez_Rodriguez_2017}. The process of manipulating the states of qubits by arbitrarily changing the gate parameters for the desired result is closely related to the training process of machine learning algorithms. For solving any specific problem, QML algorithms can be designed as a quantum circuit with a sequence of different quantum gate operations \cite{e25020287}\cite{heese2025explainingquantumcircuitsshapley}. However, Noisy Intermediate Scale Quantum (NISQ) computers are limited in resources and subject to sources of error, such as noise induced by each quantum operation \cite{Preskill_2018}\cite{Nielsen_Chuang_2010}. This makes it difficult to develop QML algorithms that perform as well as or better than conventional ones. Hence, developing QML algorithms with good performance, despite today's limited resources, has become a major challenge. To achieve this, it is important to look at several aspects of a QML task, such that the data encoding part.

Quantum encoding involves the conversion of classical information into quantum states, enabling QML algorithms to operate efficiently. This process is fundamental, as the way in which data is encoded can significantly influence the results obtained by learning models \cite{Schuld_20211}\cite{Caro_20211}. Various encoding methods have been proposed in the literature \cite{math12213318}\cite{rath2023quantumdataencodingcomparative}\cite{munikote2024comparingquantumencodingtechniques}, each with its own characteristics, advantages and disadvantages. For example, some techniques may better capture the data structure, while others may offer lower computational complexity. This strategic choice is even more relevant in the context of real-world applications where accuracy and speed are essential. Consequently, the choice of the encoding method is crucial to the success of QML algorithms.

\begin{figure}[htpb]
\centering
\scalebox{0.7}{
\begin{quantikz}
\lstick{$(x_1) \ket{0}$} & \gategroup[6,steps=3,style={dashed}, label style = {yshift=0.3cm}]{Embedding} & \gate[6]{U(x)} & & & \gate{R_y(\theta_0)} \gategroup[6,steps=6,style={dashed},label style = {yshift=0.3cm}]{Classification } &\targ{} & \ctrl{1} & & & & \\ 
\lstick{$(x_2) \ket{0}$} & & & & & \gate{R_y(\theta_1)} & & \targ{} & \ctrl{1} & & & \\
\lstick{$(x_3) \ket{0}$} & & & & & \gate{R_y(\theta_2)} & & &\targ{} & & & \\
\cdots &\cdots &\cdots &\cdots &\cdots &\cdots  & \cdots &\cdots &\cdots &\cdots &\cdots &\cdots  \\
\lstick{$(x_{n-1}) \ket{0}$} & & & & & \gate{R_y(\theta_{n-1})} &  & & & \ctrl{1} & & \\
\lstick{$(x_n) \ket{0}$} & & & & & \gate{R_y(\theta_n)} & \ctrl{-5} & & & \targ{}& & & \gate{\Z} \gategroup[1,steps=2,style={inner ysep=11pt, dashed}, label style = {yshift=0.3cm}]{\footnotesize Measurement}   & \meter{\langle \sigma_z \rangle} &  
\end{quantikz}
}
\caption{\textit{\textbf{A layer of the QML model used to carry out our simulations and the measurement part.} Only the embedding part varies. The classification and measurement parts are the same for all the encodings tested. We use data re-uploading. Thus, for each model formed with a given encoding, one layer of the model contains the encoding part and the classification part.}}
\label{fig:single model}
\end{figure}
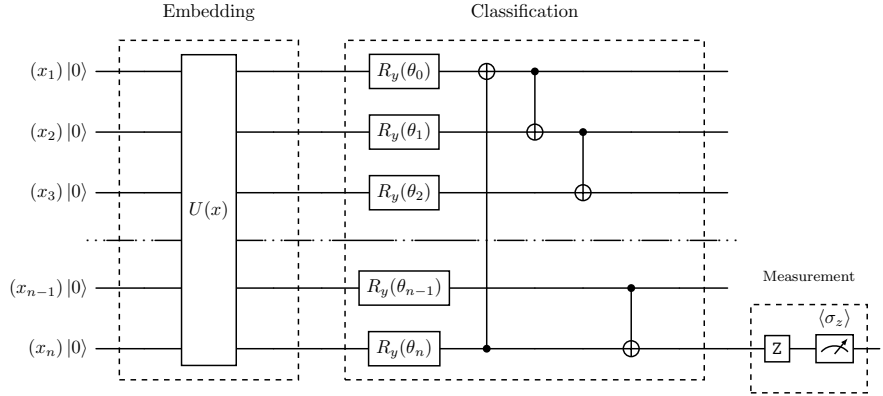

With a view to making a contribution to the choice of an appropriate encoding for a specific dataset, in this work we make a comparative study of encoding methods that are recurrent in the literature. To this end, we have used a quantum neural network classification model, presented in \autoref{fig:single model}, and trained it on real datasets, noting the results obtained for each encoding. The rest of the paper is divided in four parts. The encoding methods used for our benchmarking are presented in \autoref{Sec:encoding}, and \autoref{sec:datasets} covers the datasets used, with subsections on presentation of individual datasets, preprocessing on each dataset and details of implementation. The results for each dataset are presented in \autoref{sec:results}, with comments. We end with a conclusion and outlook (\autoref{sec:conclusion}).

\counterwithin{figure}{section}

\section{Encoding methods}\label{Sec:encoding}
\subsection{Angle encoding}
We carried out our simulations on a single classification model (\autoref{fig:single model}) on five feature maps. Three of these feature maps are different types of angle encodings \cite{PhysRevLett}\cite{Skolik_2021}. Angle encoding is a feature map in which the inputs of the data point, $x$, are fed into the circuit as rotation gate angles (usually $\RX$ and $\RY$) on the respective qubit. For a classical dataset $D = (x^d , y^d )_{d=1}^{m}$, where $m$ is the number of samples, $x^d \in \R^N$ is a $N$-dimensional input data (the features) and $y^d$ the class label for the corresponding data, we can encode $x^d$ into the quantum state
\begin{equation}\label{eq:classical data to quantum data}
    \ket{\psi^d} = \bigotimes_{j=1}^N \ket{\psi_j^d}.
\end{equation}
In the following, we will simply note $\ket{\psi^d}$ by $\ket{\psi}$.

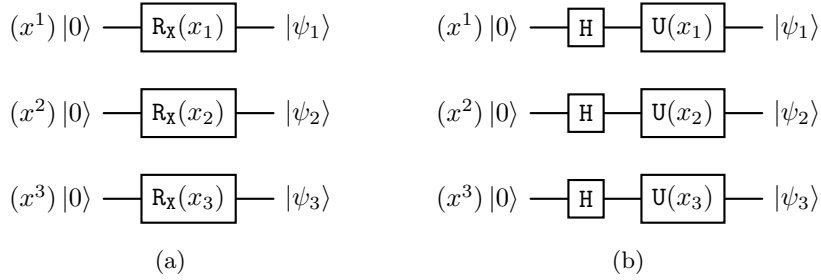
\begin{figure}[H]
    \centering
    \begin{subfigure}[t]{0.48\textwidth}
        \centering
        %\scalebox{0.99}{
  \begin{quantikz}
  \lstick{$(x^1) \ket{0}$} & \gate{\RX (x_1)} &  \rstick{$\ket{\psi_1}$}  \\
 \lstick{$(x^2) \ket{0}$} & \gate{\RX (x_2)} & \rstick{$\ket{\psi_2}$}\\
 \lstick{$(x^3) \ket{0}$} & \gate{\RX (x_3)} & \rstick{$\ket{\psi_3}$}
  \end{quantikz}
  %}
        \caption{}
        \label{fig:Simple Angle}
    \end{subfigure}%
    ~ 
    \begin{subfigure}[t]{0.48\textwidth}
        \centering
        %\scalebox{0.99}{
 \begin{quantikz}
  \lstick{$(x^1) \ket{0}$} & \gate{\HH} & \gate{\mathtt{U}(x_1)} &  \rstick{$\ket{\psi_1}$}  \\
 \lstick{$(x^2) \ket{0}$} & \gate{\HH} & \gate{\mathtt{U}(x_2)} & \rstick{$\ket{\psi_2}$}\\
 \lstick{$(x^3) \ket{0}$} & \gate{\HH} & \gate{\mathtt{U}(x_3)} & \rstick{$\ket{\psi_3}$}
  \end{quantikz}
  %}
        \caption{}
        \label{fig: pi/4 angle encoding}
    \end{subfigure}
    \caption{\textit{Simple Angle encoding (a) and $\frac{\pi}{4}$-Angle encoding (b)}.}
\end{figure}

The state $\ket{\psi_j}$ in \autoref{eq:classical data to quantum data} is given by,
\begin{equation}\label{eq:Simple Angle}
\begin{cases}
\ket{\psi_j} = \RX(x_j) \ket{0} =  \cos{\frac{x_j}{2}} \ket{0} -i  \sin{\frac{x_j}{2}} \ket{1}, \ \text{or} \\ \\
\ket{\psi_j} = \RY(x_j) \ket{0} =  \cos{\frac{x_j}{2}} \ket{0} +  \sin{\frac{x_j}{2}} \ket{1},
    \end{cases}
\end{equation}
depending on whether we use $\RX$ or $\RY$ defined by
\begin{align}
  \RX (x_j) = \begin{pmatrix}
        \cos{\frac{x_j}{2}} & -i \sin{\frac{x_j}{2}} \\ \\  -i\sin{\frac{x_j}{2}} & \cos{\frac{x_j}{2}}
    \end{pmatrix}, && \RY (x_j) = \begin{pmatrix}
        \cos{\frac{x_j}{2}} & -\sin{\frac{x_j}{2}} \\ \\ \sin{\frac{x_j}{2}} & \cos{\frac{x_j}{2}}
    \end{pmatrix}.  
    \end{align}
This encoding is called \textit{\textbf{\enquote{Simple Angle encoding}}} in the following, and the one used for our simulations is with the $\RX$ rotation gate, illustrated in \autoref{fig:Simple Angle}.

In \cite{2021QuIP...20..119C}, Chalumuri and al. propose an angle encoding (\autoref{fig: pi/4 angle encoding}) which uses \textit{Walsh-Hadamard} operator and an unitary operation with the following square matrix,
\begin{equation}\label{eq:un op}
    \mathtt{U}(x) = \begin{pmatrix}
    \cos{\big(\frac{\pi}{4} - x \big)} & \sin{\big(\frac{\pi}{4} - x\big)} \\ \\ 
    -\sin{\big(\frac{\pi}{4} - x\big)} &\cos{\big(\frac{\pi}{4} - x\big)}
    \end{pmatrix}.
\end{equation}
This encoding, called the \textbf{\textit{\enquote{$\frac{\pi}{4}$-Angle encoding}}} in the following, encodes data into the quantum state $\ket{\psi_j}$ as follows
\begin{equation}
\ket{\psi_j} = \mathtt{U}({x_j}) \HH \ket{0}.
\end{equation}
Step by step, we have
\begin{align*}
    \HH \ket{0} &= \frac{1}{\sqrt{2}} \begin{pmatrix}
     1 & 1 \\ 1 & -1   
    \end{pmatrix} \begin{pmatrix}
        1 \\ 0
    \end{pmatrix} = \frac{1}{\sqrt{2}} \begin{pmatrix}
        1 \\ 1
    \end{pmatrix}, \\ \\
    \mathtt{U}(x_j) \HH \ket{0} &= \frac{1}{\sqrt{2}} \begin{pmatrix}
    \cos{\big(\frac{\pi}{4} - x_j \big)} & \sin{\big(\frac{\pi}{4} - x_j \big)} \\ \\ 
    -\sin{\big(\frac{\pi}{4} - x_j \big)} &\cos{\big(\frac{\pi}{4} - x_j \big)}
    \end{pmatrix} \begin{pmatrix}
        1 \\ \\ 1
    \end{pmatrix} = \begin{pmatrix}
        \cos{x_j} \\ \\  \sin{x_j}
    \end{pmatrix}.
\end{align*}
Finally, the encoding state is
\begin{equation}
\ket{\psi_j} =  \cos{x_j} \ket{0} + \sin{x_j} \ket{1}.
\end{equation}

When comparing these two last angle encodings, for an input $x_j$, the quantum state $\ket{\psi_1}$ after \textit{Simple Angle encoding} using a $\RY$ rotation, and the state $\ket{\psi_2}$ after $\frac{\pi}{4}$ \textit{angle encoding} are given in equations \ref{eq:comp Simple Angle and chalumuri},
\begin{equation}\label{eq:comp Simple Angle and chalumuri}
 \ket{\psi_1} = \begin{pmatrix} \cos{\frac{x_j}{2}} \\ \\ \sin{\frac{x_j}{2}} \end{pmatrix}, \ \ \ket{\psi_2} = 
 \begin{pmatrix}\cos{x_j} \\ \\ \sin{x_j}\end{pmatrix}.
\end{equation}
This shows that the transition from the first encoding to the second would be done simply by multiplying the $x_j$ data in the first encoding by $2$.

\begin{figure}[H]
\centering
        \scalebox{1.05}{
  \begin{quantikz}
  \lstick{$(x^1) \ket{0}$} & \gate{\HH} & \gate{\RY (x_1)} & \ctrl{1} & & \targ{}& \rstick{$\ket{\psi_1}$} \\
 \lstick{$(x^2) \ket{0}$} & \gate{\HH} & \gate{\RY (x_2)} & \targ{} & \ctrl{1} & &  \rstick{$\ket{\psi_2}$} \\
 \lstick{$(x^3) \ket{0}$} & \gate{\HH} & \gate{\RY (x_3)} & & \targ{} & \ctrl{-2}  & \rstick{$\ket{\psi_3}$}
  \end{quantikz}
  }
        \caption{\textit{Entangled Angle encoding.}}
    \label{fig:Entangled angle}
\end{figure}
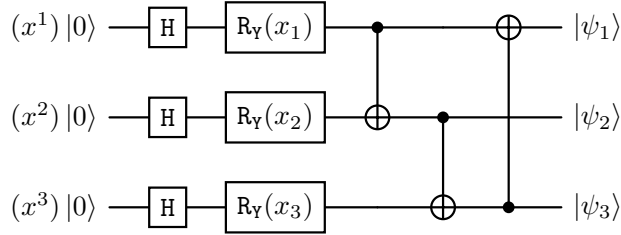

Another extension of angle encoding used for our simulations is the so called \textbf{\textit{Entangled Angle encoding}} (\autoref{fig:Entangled angle}), built on the \textbf{Simple Angle encoding} strategy by further leveraging quantum
mechanics through the use of entanglement \cite{deluca2024empiricalpowerquantumencoding}. The embedding circuit is constructed by applying a \textit{Hadamard gate} to each qubit, followed by the typical Simple Angle encoding, and a \textit{Controlled-NOT} gate between each pair of adjacent qubits. The Simple Angle encoding used here is the one with $\RY$ gate.
In all three cases of angle encoding, the number $\mathbf{n}$ \textbf{of qubits} required to encode \textbf{data of size} $\mathbf{N}$ is such that $\mathbf{n=N}$. 

\subsection{Amplitude encoding}
Amplitude encoding embeds classical features of the form $\sum_{i=1} ^N \big|x_i \big|^2 = 1$,
in an amplitude vector of a n-qubits quantum state $\ket{\psi}$, such as
\begin{align}
\ket{\psi} = \sum_{i=1} ^N x_i \ket{i},
\end{align}
where $N = 2^n$, $\ket{i}$ is the i-th computational basis state, and $x_i$ is the i-th element of $x$.
Among the methods used to carry out this encoding, we used the method developed in \cite{araujo} called the \textit{divide-and-conquer method}. This method can also be found in \cite{kerenidis2016quantumrecommendationsystems, Low_2024}, where the authors use it for the encoding of classical data into quantum data in contexts other than QML.

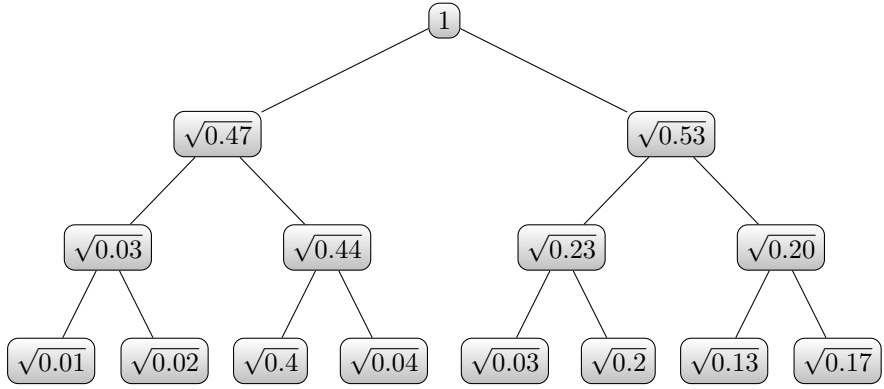
\begin{figure}[htp]
\centering
\begin{tikzpicture}[level 1/.style = {sibling distance = 6cm}, level 2/.style = {sibling distance = 2.9cm}, level 3/.style = {sibling distance = 1.5cm},
  every node/.style = {shape=rectangle, rounded corners,
    draw, align=center,
    top color=white, bottom color=gray!50}]]
\node {1}
    child { node {$\sqrt{0.47}$}
      child { node {$\sqrt{0.03}$}
      child {node{$\sqrt{0.01}$}}
      child {node{$\sqrt{0.02}$}}
      }
      child { node {$\sqrt{0.44}$}
       child { node {$\sqrt{0.4}$}}
       child { node {$\sqrt{0.04}$} }
      }
    }   
    child { node {$\sqrt{0.53}$}
      child { node {$\sqrt{0.23}$}
     child {node{$\sqrt{0.03}$}}
     child {node{$\sqrt{0.2}$}}
      }
      child { node {$\sqrt{0.20}$}
      child {node{$\sqrt{0.13}$}}
     child {node {$\sqrt{0.17}$} }
      }
    };
\end{tikzpicture}
\caption{\textit{Binary tree decomposition of our example.}}
\label{fig:binary tree}
\end{figure}

For an input $x$ of dimension $N$, the state $\ket{\psi}$ is encoded using a quantum circuit of $n$ qubits, composed of a set of $\RY$ controlled rotation gates, whose angles are found from a binary tree data. This is a tree of depth $n$, having at its root the sum of the squares of the Amplitudes $x_i$ (which is equal to 1), and at its leaves the Amplitudes $x_i$. Each intermediate node stores the sum of the squares of the Amplitudes of two child nodes. For example, for a given data
\begin{align}\label{eq:ex ampl encod}
    x^d = \big( \sqrt{0.01}, \sqrt{0.02}, \sqrt{0.4}, \sqrt{0.04}, \sqrt{0.03}, \sqrt{0.2}, \sqrt{0.13}, \sqrt{0.17} \big)
\end{align}
the binary tree is given in \autoref{fig:binary tree}.

The encoding circuit and the rotation angles are generated by \textit{Algorithm 1} from \cite{araujo}, which we encourage the reader to take a look at. Indeed, it is a two-part algorithm, the first of which computes the encoding rotation angles. The principle is as follows: \textit{starting from the input vector $x$ of dimension $N$, a new vector $x_1$ of dimension $\frac{N}{2}$ is created such that}
\begin{equation}
    x_1 = \Big( \sqrt{\big|x[2k]^2 + x[2k+1]\big|^2} \Big)_{k=0}^{\frac{N}{2}}.
\end{equation}
\textit{As long as the dimension of the new vector $x_1$ is greater than 1, $x$ is assigned $x_1$ and the procedure is repeated}. \textit{Then, for each new vector $x_1$, the rotation angles are calculated as follows}:
\begin{equation}
 \alpha =  \begin{cases}
  2 \arcsin{\Big(\frac{x[2k+1]}{x_1[k]}\Big)}, \ \ \ \text{if} \ \ x_1[k] > 0  \\
  2\pi - 2 \arcsin{\Big(\frac{x[2k+1]}{x_1[k]}\Big)}, \ \ \ \text{else}.
 \end{cases}   
\end{equation}
Using this algorithm, the output angles of our example (\autoref{fig:binary tree}) are $(1.63, 2.63,$ $1.70, 1.91,$ $0.61, 2.40, 1.70)$. 

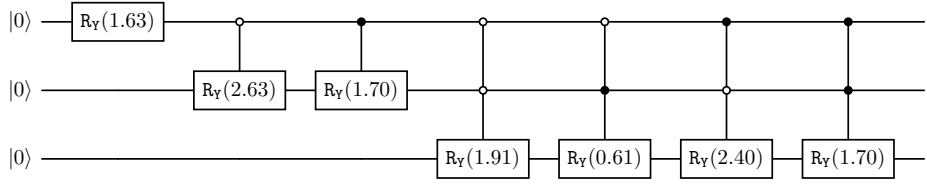
\begin{figure}[H]
    \centering 
    \scalebox{0.80}{
\begin{quantikz}
  \lstick{$\ket{0}$} & \gate{\RY (1.63)} & \octrl{1} & \ctrl{1} & \octrl{1} & \octrl{1} & \ctrl{1} & \ctrl{1} & \\  
 \lstick{$\ket{0}$} & &  \gate{\RY (2.63)} & \gate{\RY (1.70)}& \octrl{1} & \ctrl{1} & \octrl{1} & \ctrl{1} & \\ 
 \lstick{$\ket{0}$} & & & & \gate{\RY (1.91)} & \gate{\RY (0.61)} & \gate{\RY (2.40)} & \gate{\RY (1.70)} &
  \end{quantikz}
  }
    \caption{\textit{Amplitude encoding with $N=8$.}}
    \label{fig:Amplitude encoding}
\end{figure}

The second part of this algorithm generates the encoder circuit. We have not gone into details about this part. The circuit encoding the data $ x_d $ of our example in \autoref{eq:ex ampl encod} is illustrated in \autoref{fig:Amplitude encoding}.

With n qubits, we can encode $N$ data points, such that
\begin{equation}
    \begin{cases}
        2^{n-1} \leq N \leq 2^n, \ \  \text{if} \ n=1, \\
       2^{n-1} < N \leq 2^n, \ \  \text{if} \ n > 1.
    \end{cases}
\end{equation}

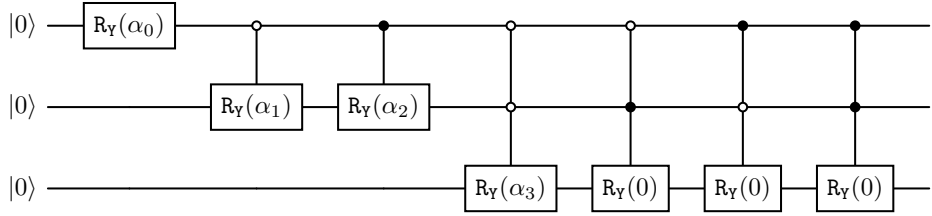
\begin{figure}[htp]
\centering 
    \scalebox{0.95}{
\begin{quantikz}
  \lstick{$\ket{0}$} & \gate{\RY (\alpha_0)} & \octrl{1} & \ctrl{1} & \octrl{1} & \octrl{1} & \ctrl{1} & \ctrl{1} & \\  
 \lstick{$\ket{0}$} & &  \gate{\RY (\alpha_1)} & \gate{\RY (\alpha_2)}& \octrl{1} & \ctrl{1} & \octrl{1} & \ctrl{1} & \\ 
 \lstick{$\ket{0}$} & & & & \gate{\RY (\alpha_3)} & \gate{\RY (0)} & \gate{\RY (0)} & \gate{\RY (0)} &
  \end{quantikz}
  }  
    \caption{\textit{Amplitude encoding with $N=6$.}}
    \label{fig:Amplitude encoding_with N=6}
\end{figure}

To show the difference between the $2^{n-1}<N<2^n$ and $N=2^n$ cases, the \autoref{fig:Amplitude encoding_with N=6} shows the case where $N=6$. This case, which requires $n=3$ qubits like the $N=8$ case, uses fewer controlled rotation gates and generates 4 rotation angles $(\alpha_0, \alpha_1, \alpha_2, \alpha_3)$ with \textit{Algorithm 1} of \cite{araujo}. So, the remaining 3 gates are set up with 0 as parameter.

\subsection{Instantaneous Quantum Polynomial-time (IQP) embedding}
 The instantaneous Quantum Polynomial-time (IQP) embedding encodes the features into qubits using diagonal gates of an IQP  circuit \cite{Havl_ek_2019}\cite{Bremner_2010}. Indeed, an IQP circuit comprises Hadamard gates $\HH$, $\RZ$ gates and the two-qubits $\mathtt{R_{ZZ}}$ for entanglement that encode higher-order data, defined in \autoref{eq:RZ and RZZ}. The $\mathtt{R_{ZZ}}$ gate provide interactions between two qubits that encode the product of two features. 
 \begin{align}\label{eq:RZ and RZZ}
 \RZ(x_i) &= \exp{\big(-i\frac{x_i}{2}\Z \big)}, &&  \mathtt{RZZ}(x_i x_j) = \exp{\big(-i\frac{x_ix_j}{2} \Z \otimes \Z \big)}.
 \end{align}

IQP circuit was originally introduced by \cite{Shepherd_2009} as a response to the difficulty of experimentally demonstrating a quantum advantage without having a complete universal computation. Indeed, in the standard circuit model, a universal set of quantum gates is used to generate the special unitary group $SU(2^n)$, which is non-Abelian. Hence, the order in which the gates are applied is crucial and reflects the temporal complexity of the computation.  IQP circuits are limited to Hamiltonians that commute with each other, thus generating an abelian subgroup of the unitary group. This commutative property means that all gates can be seen as applied simultaneously, hence the term \textit{instantaneous}. Although mathematically much more limited than universal computation, these circuits generate probability distributions that are believed to be classically inapproximable. This makes it possible to design Alice and Bob type protocols, where Alice encodes a hidden structure in a problem and Bob must, using his IQP device, produce samples that respect a specific correlation that no classical computer can reproduce efficiently. 

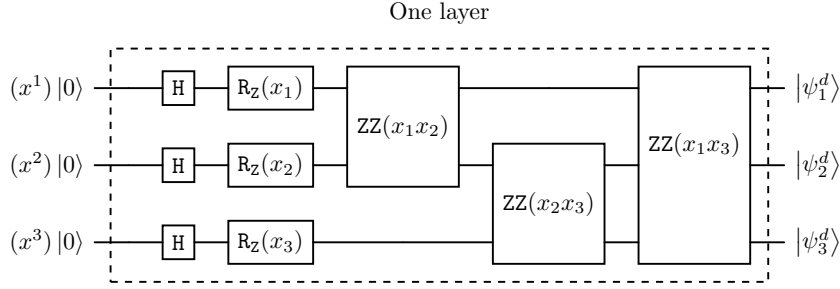
\begin{figure}[htp]
        \centering
        \scalebox{0.9}{
  \begin{quantikz}[transparent]
  \lstick{$(x^1) \ket{0}$} & \gategroup[3,steps=6,style={dashed}, label style = {yshift=0.3cm}]{One layer} & \gate{\HH} & \gate{\RZ (x_1)} & \gate[2]{\Z \Z (x_1x_2)} &  & \gate[3, label
style={yshift=0.3cm}]{\Z \Z(x_1x_3)} & \rstick{$\ket{\psi_1^d}$}  \\
 \lstick{$(x^2) \ket{0}$} & & \gate{\HH} & \gate{\RZ (x_2)} & & \gate[2]{\Z \Z (x_2 x_3)} &  &  \rstick{$\ket{\psi_2^d}$} \\
 \lstick{$(x^3) \ket{0}$} & & \gate{\HH} & \gate{\RZ (x_3)} & &  &  & \rstick{$\ket{\psi_3^d}$}
  \end{quantikz}
  }
\caption{\textit{An IQP circuit on 3-qubits: the IQP embedding is composed of at least two layers of this IQP circuit.}}
    \label{fig:IQP embedding}
\end{figure}

An IQP Embedding, as presented in \cite{autoqml} and \cite{articleHong}, takes as its input the $\ket{0}^{\otimes n}$ state, and is composed of at least two layers of the IQP circuit configuration, as presented in  \autoref{fig:IQP embedding}. 

%the name \enquote{instantaneous} comes from the fact that all the diagonal gates commute, so that there is no time order encoded into the circuit. Or equivalently, their Hamiltonians can simply be additively combined, corresponding to simultaneous evolution. \enquote{Polynomial-time} means that the process is bound to consume at most a polynomial amount. 

This method is capable of embedding $N$ data points $\big(x_1, x_2, \cdots, x_j\big)$ into $n=N$ qubits such that
\begin{equation}
    \ket{\psi} = \mathtt{U_Z} \  \HH^{\otimes n} \ \mathtt{U_Z} \  \HH^{\otimes n} \ \ket{0}^{\otimes n},
\end{equation}
where 
\begin{equation}
    \mathtt{U_Z} = \exp{\Bigg(\sum\nolimits_{\substack{j=1\\k=1\\j\ne k}}^n -i\frac{x_jx_k}{2} \Z_j \Z_k + \sum\nolimits_{j=1}^n -i\frac{x_j}{2} \Z_j\Bigg)},
\end{equation}
and $\Z_i$ is the Pauli $\Z$ applied to the $i^{\text{th}}$-qubit.

\section{Datasets and preprocessing}\label{sec:datasets}
\subsection{Datasets}

Our simulations and experiments were carried out on binary classifications, and used four benchmark datasets:
\begin{enumerate}

\item the \textbf{\textit{Malware dataset}}, a binary dataset consisting of  malicious and benign Portable Executable (PE) files from datasets Bodmas \cite{9474321} and PE Malware Machine Learning \cite{PEML}. Bodmas consists of 134,435 files. The 57,293 malicious files are fairly recent (collected between 2019 and 2020), and labeled by category (Trojan, Worm, Ransomware, $\cdots$). As for the PE Malware Machine Learning dataset, 114,737 files are malicious and come from sources such as VirusShare\footnote{\href{https://virusshare.com/}{VirusShare}} and TheZoo\footnote{\href{https://github.com/ytisf/theZoo}{TheZoo}}. The benign files (86,812) come from different versions of the Windows operating system (version $\geq$ Windows 7), for a total number of 200,549 files. We chose this dataset for our benchmarking because it corresponds to one of the key areas of our work: \textit{cybersecurity}. Specifically, because of the enormous size of the data and the fact that it is not easy to classify, this dataset offers a valuable application for quantum classification tasks \cite{bermejo2024quantumconvolutionalneuralnetworks}, and malware detection is an important research field in cybersecurity. 

\item The \textit{\textbf{Wisconsin Diagnostic Breast Cancer (\textit{WDBC}) dataset}}, originally introduced by Street et al. \cite{breast} is a binary dataset for the problem of identifying breast cancer. The dataset comprises $569$ total instances, which include $357$ benign cases and $212$ malignant ones. Each case is characterized by ten real-valued attributes pertaining to cell nuclei, such as radius, texture, perimeter, area, smoothness, compactness, concavity, concave points, symmetry, and fractal dimension. Furthermore, for each image, the mean, standard error, and the maximum values (calculated as the average of the three highest values) for these features were determined, leading to a total of 30 distinct attributes.

\item The \textit{\textbf{Modified National Institute of Standards and Technology (MNIST) dataset}} \cite{app9153169}, which consists of $60,000$ training images and $10,000$ testing images, black and white, of handwritten digits, normalized centered with $28$ pixels on each side. The database contains 10 classes, where each class refers to a digit. In order to carry out our simulations focused on binary classifications, we first used the $0$ and $1$ classes, and then $0$ and $8$ in a second stage in order to have two binary datasets. In the following, we refer to them as \textbf{\textit{MNIST 01 dataset}} (12665 training samples and 2115 test samples) and \textbf{\textit{MNIST 08 dataset}} (11774 training samples and 1954 test samples), respectively.
The choice of these classes is motivated by the fact that while 0 and 1 have distinct characteristics that can make them easier to differentiate, the distinction between $0$ and $8$ presents a moderate challenge, which can be interesting for testing the robustness of the models without being too complex. The choice of WDBC and MNIST datasets is motivated by their popularity.

\end{enumerate}

\subsection{Preprocessing}\label{subsec:preprocessing}

We extracted from each dataset the 8 features with the highest impact and the lowest correlation with each other. With the WDBC dataset, we used the \textit{Permutation feature importance} metric from \textit{Scikit-learn}\footnote{\href{https://scikit-learn.org/stable/modules/permutation_importance.html}{Permutation feature importance}} to extract these features. The work involved in ranking the most impacting features of the Malware dataset is developed in \textit{Benjamin Marais}'s thesis \cite{marais:tel-04416984}. 

Data from WDBC and Malware datasets were rescaled in $\big[0, \frac{\pi}{2}\big]$. The image data from MNIST was resized using the \textit{Principal Component Analysis (PCA)} \cite{jolliffe} to suitable dimensions, and then also rescaled in $\big[0, \frac{\pi}{2}\big]$. 

For the training of MNIST and Malware datasets, we used \textbf{4000} samples. However, regarding tests, \textbf{2000} samples are used for Malware and MNIST01 datasets, while for MNIST08, all the \textbf{1954} test samples are used. Regarding the WDBC dataset, we divided the dataset into a training set and a testing set. We used a common split ratio of \textbf{80:20}, where 80\% of the data $(455)$ is used for the training and the rest $(114)$ for evaluating the model’s performance.
 
\subsection{Implementation}
For our simulations, we used \textbf{Qiskit} simulators from \textit{IBM} \cite{quantumcomputingqiskit}. The architecture of our layers of the \autoref{fig:single model} in which the data is reuploaded at the beginning of each layer is justified by the fact that this reuploading gives better results than when there is no reuploading \cite{Perez_Salinas_2020}.
We trained over 5 epochs. 

For each of the 4 datasets, we carried out simulations on the 5 encodings for $\mathbf{N}$ \textbf{ features} and $\mathbf{M}$ \textbf{ layers}. First, we chose the number of features and layers as a power of 2. Then we looked at the behavior of Amplitude encoding in cases where $\mathbf{N}$ is not a power of 2 (\autoref{fig:Amplitude encoding_with N=6}). This is how we set
\begin{align}\label{eq:features and layers}
    N \in \big\{4,6,8\big\}, && M \in \big\{2, 4\big\}.
\end{align}

Depending on the number of \textbf{features} $\mathbf{N}$, the number of \textbf{qubits} $\mathbf{n}$ used for each encoding is given in \autoref{tab:qubits_number}.
\begin{table}[ht]
    \centering
    \begin{tabular}{|c|c|}
    \hline
      Encoding & Number of qubits \\
      \hline
       Angle encoding  & $n=N$ \\
       \hline
       Amplitude encoding & $n\geq \log_2(N), \ n\in \N^*$ \\
       \hline 
       IQP encoding  & $n=N$ \\
       \hline
    \end{tabular}
     \caption{\textit{Number of qubits according to the encoding used.}}
    \label{tab:qubits_number}
\end{table}

We looked at training accuracy over \textbf{5 epochs} with a learning rate $\mathbf{\eta =0.1}$, test accuracy at the end of training, and the F1 score. We present these results in the next section.

\section{Results}\label{sec:results}
For each dataset, we present the performance of the model of \autoref{fig:single model}, according to each encoding method and the parameters given in \ref{eq:features and layers}. In this section, these results (train accuracies, test accuracies and F1-scores) are presented in tabular form. For a better evaluation of the evolution of performance, we present in the Appendix (\autoref{sec:appendix}), a graphical representation of the precision of the training according to each epoch.

\subsection{Malware dataset}

The results obtained with the Malware dataset are summarized in \autoref{tab:malware1}, and \autoref{tab:malware2}. The highest F1-score is $90\%$, obtained with Amplitude encoding on 8 features, 4 layers. This encoding gives the best accuracies for models where the number of features is a power of 2 (with F1-scores varying between $88\%$ and $90\%$). However, we note that with 6 features, the model does not perform as well. With 4 layers, we do not have the best accuracies, and with 2 layers, the model has the same training accuracy ($83\%$) on all epochs. With 6 features, the form of this encoding, as shown in the figure, uses 3 qubits as with 8 features. However, unlike the latter case, where 4 consecutive rotation gates are applied to the last qubit, the last qubit with 6 features applies only one rotation gate, the other three becoming an identity transformation due to the zero angles. This could explain the difference in results between 8 and 6 features. 

The lowest F1-score ($82.11\%$) comes from the $\frac{\pi}{4}$-Angle encoding, obtained with 4 features and 2 layers. Further on, we note that this encoding gives the lowest precision on all 2-layers models. This is not the case with 4 layers, where this encoding gives the best accuracies with 6 features and 4 layers (train of $>88\%$, test of $90\%$ and an F1-score of $89,26\%$). We can deduce that the more layers the model has, the better it performs. We also note that results with 8 features are no better than those with 6 features, whatever the number of layers. Knowing that features are ranked in order of importance, this could be explained by the fact that 6 features are enough for us to get good accuracies with $\frac{\pi}{4}$-Angle encoding. Increasing the number of features would add nothing to the model, and would rather seem to worsen the results. This means the more layers the model has, the better it performs. It is an advantage in that, in the current era of Noisy Intermediate-Scale Quantum (NISQ), QML algorithms are developed and tested with very few resources (few qubits).

The same observations are made with IQP encoding, where with the same number of features, 4-layers models perform better than 2-layers ones. However, in terms of number of features, with just 4 features,  IQP achieves its best train and test accuracies (with a F1-score $> 89\%$).  Increasing the number of features does not seem to improve results. We also note that with 4 features, training accuracies with 2 layers stagnate at the same value from the first epoch onwards, which is not the case with 4 layers, where the model starts from a training accuracy that improves with each epoch. This encoding would need few features (thus few qubits) and a number of layers greater than 2 to establish a good relationship between features and target. One interpretation of this observation would be that the IQP needs more parameters to improve with epochs.

\renewcommand{\arraystretch}{2.2} %donne la distance entre les lignes%

\begin{table}[H]
    \centering
    \resizebox{\columnwidth}{!}{
    \begin{tabular}{|c|c|c|c|c|c|c|c|c|} 
       \hline
       \small 
\textbf{Models} & \textbf{Encodings} & \textbf{ep1} & \textbf{ep2} & \textbf{ep3} & \textbf{ep4} & \textbf{ep5} & \textbf{test} & \textbf{F1-score}\\
       \hline
        \multirow{5}{*}{\textbf{(4F, 4L)}} & {Simple Angle} & 0.8568& 0.8615& 0.8602& 0.8612& 0.8618& 0.877& 0.8806 \\ \cline{2-9} 
       & {$\frac{\pi}{4}$ {angle}} & 0.8525& 0.8755& 0.8702& 0.8730& 0.8712& 0.8825& 0.8724 \\ \cline{2-9} 
        & {Entangled Angle} & 0.8572& 0.8672& 0.8740& 0.8745& 0.8695& 0.8305& 0.8439 \\ \cline{2-9}
        & \textbf{Amplitude} & 0.8452& 0.8670& 0.8682& 0.8695& 0.8695& 0.8945& \textbf{0.8910} \\ \cline{2-9}
        & \textbf{IQP} & 0.8785& 0.8890& 0.8912& 0.8922& 0.8932& 0.898& \textbf{0.8914} \\ \cline{2-9}
       \Xhline{3\arrayrulewidth}       
       \multirow{5}{*}{\textbf{(4F, 2L)}} & {Simple Angle} & 0.7993& 0.8000& 0.8035& 0.8027& 0.8020& 0.8155& 0.8305\\ \cline{2-9} 
       & { $\frac{\pi}{4}$-Angle} & 0.8020& 0.8040& 0.7927& 0.7913& 0.7905& 0.806 & 0.8211 \\ \cline{2-9} 
        & {{Entangled Angle}} & 0.7980& 0.8245& 0.8270& 0.8255& 0.8245& 0.8205& 0.8362 \\ \cline{2-9}
        & \textbf{Amplitude} & 0.8562& 0.8642& 0.8522& 0.8650& 0.8570& 0.884& \textbf{0.8844} \\ \cline{2-9}
        & {IQP} & 0.8133& 0.8175& 0.8163& 0.8180& 0.8190& 0.8255& 0.8401 \\ \cline{2-9}
       \Xhline{3\arrayrulewidth}    
       \multirow{5}{*}{\textbf{(6F, 4L)}} & {Simple Angle} & 0.8230& 0.8223& 0.8215& 0.8213& 0.8223& 0.8475& 0.8563 \\ \cline{2-9} 
       & { $\mathbf{\frac{\pi}{4}}$ \textbf{angle}} & 0.8780& 0.8790& 0.8800& 0.8828& 0.8865& 0.9005& \textbf{0.8926} \\ \cline{2-9} 
        & \textbf{Entangled Angle} & 0.8558& 0.8730& 0.8720& 0.8745& 0.8730& 0.892& \textbf{0.8913} \\ \cline{2-9}
        & {{Amplitude}} & 0.8678& 0.8748& 0.8792& 0.8770& 0.8802& 0.883& 0.8763\\ \cline{2-9}
        & {IQP} & 0.8742& 0.8850& 0.8915& 0.8880& 0.8875& 0.8935& 0.8836 \\ \cline{2-9}
       \Xhline{3\arrayrulewidth} 
       \multirow{5}{*}{\textbf{(6F, 2L)}} & {Simple Angle} & 0.7680& 0.7867& 0.7860& 0.7853& 0.7880& 0.811& 0.8288\\ \cline{2-9} 
       & { $\frac{\pi}{4}$-Angle} & 0.8035& 0.8283& 0.8267& 0.8260& 0.8235& 0.833& 0.8306\\ \cline{2-9} 
        & {{Entangled Angle}} & 0.8065& 0.8200& 0.8227& 0.8223& 0.8197& 0.822& 0.8374 \\ \cline{2-9}
        & {\textbf{Amplitude}} & 0.8335& 0.8335& 0.8327& 0.8340& 0.8317& 0.841& \textbf{0.8515} \\ \cline{2-9}
        & IQP & 0.8283& 0.8253& 0.8243& 0.8280& 0.8245& 0.8305& 0.8384 \\ \cline{2-9}
       \Xhline{3\arrayrulewidth} 
    \end{tabular}}
    \caption{\textit{The five encodings on the \textbf{Malware dataset}: training accuracies after 5 epochs, test and F1-Score accuracies for $N$ features and $M$ layers, with $N=\{4,6\}, \  M \in \big\{2, 4\big\}$}.}
\label{tab:malware1}
\end{table}

The model for which Entangled Angle encoding is better than the others is (8F, 2L). Comparing the results obtained with this encoding on the 6 models, we note that the best models are the one with 6 features and 4 layers, and the one with 8 features, 2 layers, and the worst results are with 6 features and 2 layers. There's a striking difference in results for models with 6 features (F1-score of $89\%$ with 4 features against $83.74\%$ with 2 features). The same difference appears with 8 features, where results are lower with 4 layers (train of $\approx 84\%$, test of $82.8\%$ and F1-score of $84.19\%$), than with 2 layers (train of $\approx 88\%$, test of $89.65\%$ and F1-score of $89.12\%$). This would mean that with fewer features, Entangled Angle encoding needs more layers for better accuracy. In other words, there should be a kind of balance between the number of features (the most impactful) and the number of layers for Entangled Angle encoding to return good precision.

\begin{table}[H]
    \centering
    \resizebox{\columnwidth}{!}{
    \begin{tabular}{|c|c|c|c|c|c|c|c|c|} 
       \hline
\textbf{Models} & \textbf{Encodings} & \textbf{ep1} & \textbf{ep2} & \textbf{ep3} & \textbf{ep4} & \textbf{ep5} & \textbf{test} & \textbf{F1-score}\\
       \hline
\multirow{5}{*}{\textbf{(8F, 4L)}} & {Simple Angle} & 0.8492& 0.8425 & 0.8465& 0.8390& 0.8427& 0.84857& 0.8574\\ \cline{2-9} 
       & { $\mathbf{\frac{\pi}{4}}$ \textbf{angle}} & 0.8722& 0.8802& 0.8782& 0.8808& 0.8825& 0.8885& \textbf{0.8816} \\ \cline{2-9} 
        & {Entangled Angle} & 0.8450& 0.8357& 0.8195& 0.8205& 0.8203& 0.828& 0.8419 \\ \cline{2-9}
        & {{\textbf{Amplitude}}} & 0.8782& 0.8770& 0.8810& 0.8815& 0.8752& 0.901& \textbf{0.9000} \\ \cline{2-9}
        & {IQP} & 0.8130& 0.8522& 0.8645& 0.8528& 0.8498& 0.8695& 0.8648\\ \cline{2-9}
       \Xhline{3\arrayrulewidth}
       \multirow{5}{*}{\textbf{(8F, 2L)}} & {Simple Angle} & 0.7875& 0.8047& 0.8007& 0.8027& 0.8033 & 0.8195 & 0.8353 \\ \cline{2-9} 
       & {$\frac{\pi}{4}$ {Angle}} & 0.7903& 0.7945& 0.8243& 0.8275& 0.8277 & 0.830 & 0.8281 \\ \cline{2-9} 
        & {\textbf{Entangled Angle}} & 0.8377& 0.8782& 0.8750& 0.8722& 0.8745 & 0.8965 & \textbf{0.8912} \\ \cline{2-9}
        & {\textbf{Amplitude}} & 0.8610& 0.8660& 0.8625& 0.8640& 0.8652& 0.8856 & \textbf{0.8860} \\ \cline{2-9}
        & {IQP} & 0.7913& 0.8117& 0.8107& 0.8097& 0.8110 & 0.827 & 0.8402 \\ \cline{2-9}
       \Xhline{3\arrayrulewidth}
\end{tabular}}
    \caption{\textit{The five encodings on the \textbf{Malware dataset}: training accuracies after 5 epochs, test and F1-Score accuracies for $N$ features and $M$ layers, with $N=8, \  M \in \big\{2, 4\big\}$}.}
\label{tab:malware2}
\end{table}

The Simple Angle encoding does not appear to be the best encoding on any of the six models. Of all the results obtained with this encoding, it performs better with 4 layers than with 2. We note that the best results obtained out of the 6 models with this encoding are the case with 4 features and 4 layers. Increasing the number of features also seems to decrease model performance. Since features are ranked in descending order of importance, the Simple Angle encoding applied with a $\RX$ rotation needs few features (the most important in the dataset) to make good predictions. Comparing Simple Angle encoding with the 2 other angle encodings (Entangled Angle and $\frac{\pi}{4}$-Angle) on the 6 models, Simple Angle and $\frac{\pi}{4}$-Angle perform better than Entangled Angle encoding on (4F, 4L). With 6 features and the same number of layers, the best accuracies are obtained with Entangled Angle and $\frac{\pi}{4}$-Angle encodings. The $\frac{\pi}{4}$-Angle encoding returns better accuracies than the others on (8F, 4L), while the Entangled Angle encoding is the best with 8 features and 2 layers.  

In summary, we can see that, depending on the number of features and layers, some encodings perform better than others. However, Amplitude encoding is the one that stands out with the best results, on all models except the 6-feature, 4-layers model. On the other hand, encoding by Simple Angle is limited to average results (and is never the best encoding), whatever the model. To make it easier to compare the different encodings, on each model we have highlighted in bold the encoding returning the best accuracies. We have done this with the other datasets that will follow, which will therefore make it possible to highlight the encodings with uniform results across the different datasets.

\subsection{WDBC dataset}

The behavior of each encoding on the Malware dataset may be specific to either the Malware data or the models (number of features and layers), or both. The results on the WDBC dataset are presented in \autoref{tab:WDBC1} and \autoref{tab:WDBC2}. The best results obtained have F1-scores of over $92\%$.

Amplitude encoding gives the worst results on most models, with a max F1-score of $73.78\%$, and a min of $54.76 \%$. This maximum is reached with 8 features and 4 layers. As the WDBC dataset has fewer samples than the Malware dataset, these results could be explained by the fact that Amplitude encoding in a case of very few data for training needs not only more layers, but also more features. And in reference to the Malware case, when we have a large number of samples, this encoding would require fewer layers and features. The results obtained with 6 features on WDBC should therefore in principle be better than those with 4 features. However, we note that the case (6F, 4F) gives the lowest accuracy. This could be linked to the shape of the encoding circuit (\autoref{fig:Amplitude encoding_with N=6}), where the last rotation gates have zero angles. The accuracies returned by the Amplitude encoding on WDBC data, compared to the results on the Malware data may suggest that, \textit{this encoding is not the most appropriate for a dataset with few data}.

The best results are obtained with angle encoding, where Simple Angle encoding returns the highest train and test accuracies ($> 94 \%$), and the highest F1-score $(> 92\%)$ with 4 features and 4 layers. With this encoding, there was a significant difference between the training accuracies of the different epochs. This was not the case with the Malware dataset, where not only were the worst results obtained with this encoding, but the test accuracies stagnated at a single value from the very first epochs. However, one behaviour observed on the Malware dataset that recurs with WDBC is the fact that the best accuracies with Simple Angle encoding are achieved with just 4 features. Increasing the number of features makes the results worsen. This last observation is also made with $\frac{\pi}{4}$-Angle encoding.

Simple Angle and $\frac{\pi}{4}$-Angle encodings behave in the same way with 4 features, with equal test accuracies, and F1-score measurements on the same number of layers. However, on each epoch training is better with Simple Angle encoding. This suggests that, of the two angle encodings, it is better to use the Simple Angle in cases with very few data, such as the WDBC dataset. The $\frac{\pi}{4}$-Angle encoding, on the other hand, would be a better choice than the simple one for large amounts of data (Malware dataset). As we increase the number of qubits (and therefore features), we note that with 6 qubits, a significant difference emerges from the results, where $\frac{\pi}{4}$-Angle encoding returns the best training and test accuracies, and the best F1-score. When we go up to 8 features, the same observation is made with just 4 layers. In the case of 2 layers, results for the Simple Angle improve (from $75.36\%$ to $80.48\%$), while results for the $\frac{\pi}{4}$-Angle encoding fall (from $80\%$ to $64.61\%$). This big difference between the F1-scores may be linked to the number of WDBC samples and the features ranked in order of importance. With a large number of samples, as in the Malware dataset, the 2-layers model is better able to learn from additional data. This would explain why, with 8 features, this behaviour is not observed on the Malware dataset with $\frac{\pi}{4}$-Angle encoding. Furthermore, in a framework of features ranked from most to least impacting, going up to 8 features seems to degrade the results. This explains why on WDBC, with 4 and 6 features, the difference in performance between the 4-layers and 2-layers models is less significant than with 8 features.

\renewcommand{\arraystretch}{2.2} %donne la distance entre les lignes%

\begin{table}[H]
    \centering
    \resizebox{\columnwidth}{!}{
    \begin{tabular}{|c|c|c|c|c|c|c|c|c|} 
       \hline
\textbf{Models} & \textbf{Encodings} & \textbf{ep1} & \textbf{ep2} & \textbf{ep3} & \textbf{ep4} & \textbf{ep5} & \textbf{test} & \textbf{F1-score}\\
       \hline
        \multirow{5}{*}{\textbf{(4F, 4L)}} & {\textbf{Simple Angle}} & 0.8989& 0.9275& 0.9319& 0.9407& 0.9297& 0.9493& \textbf{0.9230} \\ \cline{2-9} 
       & {$\mathbf{\frac{\pi}{4}}$ \textbf{Angle}} & 0.8747& 0.9275& 0.9297& 0.9341& 0.9231& 0.9470& \textbf{0.92} \\ \cline{2-9} 
        & {Entangled Angle} & 0.9033& 0.9275& 0.9297& 0.9297& 0.9363& 0.9385& 0.9090 \\ \cline{2-9}
        & {Amplitude} & 0.6374& 0.6703& 0.6813& 0.6857& 0.6769& 0.6140& 0.6206 \\ \cline{2-9}
        & \textbf{IQP} & 0.8505& 0.9165& 0.9253& 0.9341& 0.9341& 0.9385& \textbf{0.9176} \\ \cline{2-9}
       \Xhline{3\arrayrulewidth}       
       \multirow{5}{*}{\textbf{(4F, 2L)}} & {Simple Angle} & 0.8725& 0.9319& 0.9275& 0.9231& 0.9275& 0.9122& 0.8888 \\ \cline{2-9} 
       & {$\frac{\pi}{4}$-Angle} & 0.8571& 0.8835& 0.8769& 0.8879& 0.9011& 0.9122& 0.8809 \\ \cline{2-9} 
        & \textbf{Entangled Angle} & 0.8857& 0.9187& 0.9143& 0.9231& 0.9187& 0.9379& \textbf{0.9075} \\ \cline{2-9}
        & {Amplitude} & 0.6791& 0.6615& 0.6637& 0.6615& 0.6725& 0.6842& 0.6326 \\ \cline{2-9}
        & {IQP} & 0.7604& 0.8198& 0.8110& 0.8088& 0.8066& 0.8333& 0.7076 \\ \cline{2-9}
       \Xhline{3\arrayrulewidth}
       \multirow{5}{*}{\textbf{(6F, 4L)}} & \textbf{Simple Angle} & 0.8857& 0.9341& 0.9319& 0.9407& 0.9363& 0.9385& \textbf{0.9135} \\ \cline{2-9} 
       & {$\frac{\pi}{4}$-Angle} & 0.8462& 0.9099& 0.9187& 0.9077& 0.9121& 0.9035& 0.8571 \\ \cline{2-9} 
        & \textbf{Entangled Angle} & 0.8901& 0.9451& 0.9516& 0.9473& 0.9451& 0.9385& \textbf{0.9156} \\ \cline{2-9}
        & {Amplitude} & 0.6418& 0.6396& 0.6352& 0.6505& 0.6308& 0.6666& 0.5476\\ \cline{2-9}
        & {IQP} & 0.8198& 0.9011& 0.9077& 0.8945& 0.9055& 0.9122& 0.8684 \\ \cline{2-9}
       \Xhline{3\arrayrulewidth}       
       \multirow{5}{*}{\textbf{(6F, 2L)}} & {Simple Angle} & 0.8066& 0.7495& 0.7407& 0.7582& 0.7890& 0.8333& 0.7164 \\ \cline{2-9} 
       & {$\mathbf{\frac{\pi}{4}}$ \textbf{angle}} & 0.7385& 0.7912& 0.8484& 0.8857& 0.9055& 0.8947& \textbf{0.8461} \\ \cline{2-9} 
        & {Entangled Angle} & 0.8549& 0.8769& 0.8813& 0.8813& 0.8791& 0.8684& 0.7887 \\ \cline{2-9}
        & {Amplitude} & 0.6703& 0.6769& 0.6681& 0.6747& 0.6747& 0.7017& 0.6530 \\ \cline{2-9}
        & {IQP} & 0.7429& 0.7846& 0.8264& 0.8418& 0.8484& 0.8771& 0.8157\\ \cline{2-9}
       \Xhline{3\arrayrulewidth}

    \end{tabular}}
    \caption{\textit{The five encodings on the \textbf{WDBC dataset}: training accuracies after 5 epochs, test and F1-Score accuracies for $N$ features and $M$ layers, with $N=\{4,6\}, \  M \in \big\{2, 4\big\}$}.}
\label{tab:WDBC1}
\end{table}

\begin{table}[H]
    \centering
    \resizebox{\columnwidth}{!}{
    \begin{tabular}{|c|c|c|c|c|c|c|c|c|} 
       \hline
\textbf{Models} & \textbf{Encodings} & \textbf{ep1} & \textbf{ep2} & \textbf{ep3} & \textbf{ep4} & \textbf{ep5} & \textbf{test} & \textbf{F1-score}\\
       \hline
\multirow{5}{*}{\textbf{(8F, 4L)}} & {Simple Angle} & 0.8088& 0.8352& 0.8418& 0.8505& 0.8615& 0.8508& 0.7536 \\ \cline{2-9} 
       & { $\frac{\pi}{4}$ {angle}} & 0.7670& 0.8154& 0.8308& 0.8484& 0.8593& 0.8684& 0.8 \\ \cline{2-9} 
        & \textbf{Entangled Angle} & 0.8879& 0.9297& 0.9385& 0.9363& 0.9319& 0.9298& \textbf{0.8947} \\ \cline{2-9}
        & {{Amplitude}} & 0.6637& 0.7099& 0.7209& 0.7121& 0.7143& 0.7631& 0.7378 \\ \cline{2-9}
        & {IQP} & 0.7538& 0.8154& 0.8374& 0.8571& 0.8725& 0.8596& 0.7714\\ \cline{2-9}
       \Xhline{3\arrayrulewidth}
       \multirow{5}{*}{\textbf{(8F, 2L)}} & {Simple Angle} & 0.7780& 0.8593& 0.8615& 0.8549& 0.8527& 0.8596& 0.8048 \\ \cline{2-9}
       & {$\frac{\pi}{4}$ {angle}} & 0.7648& 0.7846& 0.7802& 0.8088& 0.7890& 0.7982& 0.6461 \\ \cline{2-9} 
        & \textbf{Entangled Angle} & 0.7736& 0.8462& 0.8879& 0.8989& 0.9033& 0.9290& \textbf{0.89}\\ \cline{2-9}
        & {{Amplitude}} & 0.6308& 0.6901& 0.7011& 0.6989& 0.7077& 0.7192& 0.6734 \\ \cline{2-9}
        & {IQP} & 0.7473& 0.8198& 0.8440& 0.8286& 0.8484& 0.8070& 0.6562\\ \cline{2-9}
       \Xhline{3\arrayrulewidth}
\end{tabular}}
    \caption{\textit{The five encodings on the \textbf{WDBC dataset}: training accuracies after 5 epochs, test and F1-Score accuracies for $N$ features and $M$ layers, with $N=8, \  M \in \big\{2, 4\big\}$}.}
\label{tab:WDBC2}
\end{table}

One encoding that particularly attracts our attention on WDBC is Entangled Angle encoding. With a number of features $N$ that is a power of 2 ($N=4, 8$), we have the same test accuracy and F1-score whatever the number of layers, with the 4-layers cases returning the best accuracies. Comparing these results with those for the Malware dataset, we realize that the hypothesis of a balance between number of features (to the power of 2) and number of layers does not seem to hold true with very few data. However, on both Malware and WDBC datasets, the best accuracies with Entangled Angle encoding are obtained with the case (6F, 4L) and the lowest with (6F, 2L). In fact, there is a significant difference in accuracies between these two cases, whether in training, testing or F1-score measurement. On the WDBC dataset, this encoding is what Amplitude encoding is on the Malware dataset. In other words, on several models, it is one of the best performers, unlike others which perform very well on some models and less well on others. This may be linked to the entanglement created between the data during encoding, and the much smaller number of samples than in the case of the Malware dataset.

IQP encoding seems to have the same behavior as with Malware dataset. However, there is a difference in accuracy between the same model on both datasets. With WDBC, IQP encoding gives good accuracies with 4 out of 4 layers and 6 features, which are the best accuracies with IQP on WDBC. These models also gave the best accuracies with the Malware dataset. With 8 features, the imbalance in the WDBC dataset (number of benign samples greater than the number of malignant samples) seems to be reflected, whatever the number of layers. Indeed, we have an F1-score of $77.14\%$ with a train accuracy of up to $87\%$ and a test accuracy of $85.96\%$ on 4 layers. On 2 layers, the F1-score is lower ($65.62\%$) with learning and test accuracies above $80\%$. Our intuition is that, with very few data and layers, the model trained with IQP encoding concentrates on learning the excess class in the dataset (benign class), neglecting the other class. This justifies good training and testing, but a low F1-scores.

At this level, we can put forward the hypothesis that Amplitude encoding is the most suitable for models trained on a large number of samples (Malware dataset). On the other hand, for a very small amount of data, as in the case of WDBC dataset, the Simple Angle and Entangled Angle encodings encode the data better for a high-performance model. In order to validate this last hypothesis, we ran the same simulations on the Malware dataset on a number of train and test data equal to those of the WDBC dataset. As for the first hypothesis, the results on the MNIST datasets presented in the next section were used to verify it.

\subsection{MNIST datasets}

 \autoref{tab:MNIST 01_1} and \autoref{tab:MNIST 01_2} present results on \textbf{MNIST 01 dataset}. \autoref{tab:MNIST 08_1} and \autoref{tab:MNIST 08_2} present results on \textbf{MNIST 08 dataset}. MNIST is a popular dataset that generally returns good results, as these results show. For a better visualization, also refer to \autoref{fig:MNIST01} and \autoref{fig:MNIST08} in the appendix. Comparing the results between the two datasets, the maximum accuracies achieved with \textbf{MNIST01} are up to $99\%$ (train and test accuracies, F1-score). Whereas with \textbf{MNIST08}, the maximum train and test accuracies achieved are $96\%$, for a maximum F1-score of $96\%$, all models combined. This is surely due by the fact that classes 0 and 8 are more difficult to differentiate than classes 0 and 1. 

Simple Angle encoding returns, on average, the lowest accuracies on the MNIST01 dataset (\autoref{fig:MNIST01}). Case (6F, 2L) is particularly poor, with train accuracies ranging from $61\% - 67\%$, and a F1-score of $73.65\%$.  These results are the worst of all MNIST datasets. Note that this encoding has the same behavior on both MNIST datasets. For the same number of features, the 4-layers cases perform better than the 2-layers cases. The model with 4 features and 4 layers is the only model for which Simple Angle encoding gives the best accuracy (on both MNIST). On MNIST01, with less than 8 features and 2 layers, the training reaches a maximum value from the first epoch. This value starts to decrease from the second epoch, as if the model were learning in the opposite direction. In general, this encoding seems unsuitable for making predictions on MNIST data (and therefore on image data).

With $\frac{\pi}{4}$-Angle encoding, the results obtained with 6 features are better than those obtained with 4 features and with 8 features. On \textbf{MNIST08}, this encoding never gives the best results, whatever the model. For $N= 4, 6$, we note a significant difference between train and test accuracies, where the train is better than the test with 4 features. On the other hand, with 6 features, the opposite behaviour is observed. On MNIST01, however, this difference is not observed. The $\frac{\pi}{4}$-Angle encoding on MNIST01 gives the best accuracies on all 2-layerss models. However, training accuracies quickly stagnate around the same training value on 2-layers models, whatever the number of features is. This is rather in line with the observation made on previous datasets, that this encoding is better with 4 layers than with 2.

\renewcommand{\arraystretch}{2.2} %donne la distance entre les lignes%

\begin{table}[H]
    \centering
    \resizebox{\columnwidth}{!}{
    \begin{tabular}{|c|c|c|c|c|c|c|c|c|} 
       \hline
       \small 
\textbf{Models} & \textbf{Encodings} & \textbf{ep1} & \textbf{ep2} & \textbf{ep3} & \textbf{ep4} & \textbf{ep5} & \textbf{test} & \textbf{F1-score}\\
       \hline
        \multirow{5}{*}{\textbf{(4F, 4L)}} & {Simple Angle} & 0.9898& 0.9948& 0.9940& 0.9942& 0.9948&0.971& 0.9722  \\ \cline{2-9} 
       & {$\frac{\pi}{4}$ {angle}} & 0.9790& 0.9840& 0.9835& 0.9782& 0.9762& 0.9705& 0.9731 \\ \cline{2-9} 
        & \textbf{Entangled Angle} & 0.9882& 0.9945& 0.9948& 0.9950& 0.9950& 0.9945& \textbf{0.9948} \\ \cline{2-9}
        & \textbf{{Amplitude}} & 0.9855& 0.9928& 0.9932& 0.9930& 0.9938& 0.997& \textbf{0.9971} \\ \cline{2-9}
        & \textbf{{IQP}} & 0.9850& 0.9928& 0.9935& 0.9938& 0.9930& 0.9945& \textbf{0.9948} \\ \cline{2-9}
       \Xhline{3\arrayrulewidth}       
       \multirow{5}{*}{\textbf{(4F, 2L)}} & {Simple Angle} & 0.9253& 0.9170& 0.8340& 0.8370& 0.8330 & 0.7855& 0.8243\\ \cline{2-9} 
       & {$\mathbf{\frac{\pi}{4}}$ \textbf{angle}} & 0.9822& 0.9855& 0.9832& 0.9842& 0.9848& 0.984& \textbf{0.9852} \\ \cline{2-9} 
        & {{Entangled Angle}} & 0.8955& 0.9133& 0.9185& 0.9143& 0.9160& 0.8845& 0.9027 \\ \cline{2-9}
        & \textbf{{Amplitude}} & 0.9825& 0.9862& 0.9875& 0.9875& 0.9875& 0.9825& \textbf{0.9839} \\ \cline{2-9}
        & {IQP} & 0.9223& 0.9390& 0.9377& 0.9373& 0.9357& 0.873& 0.8938 \\ \cline{2-9}
       \Xhline{3\arrayrulewidth}    
       \multirow{5}{*}{\textbf{(6F, 4L)}} & {Simple Angle} & 0.9610& 0.9798& 0.9808& 0.9808& 0.9798& 0.987& 0.9879 \\ \cline{2-9} 
       & {$\mathbf{\frac{\pi}{4}}$ \textbf{angle}} & 0.9838& 0.9922& 0.9912& 0.9928& 0.9920& 0.9915& \textbf{0.9921}\\ \cline{2-9} 
        & {{Entangled Angle}} & 0.9623& 0.9740& 0.9735& 0.9732& 0.9740& 0.967& 0.9701 \\ \cline{2-9}
        & \textbf{Amplitude} & 0.9905& 0.9938& 0.9938& 0.9938& 0.9942& 0.9955& \textbf{0.9958}\\ \cline{2-9}
        & {IQP} & 0.9477& 0.9475& 0.9465& 0.9485& 0.9473& 0.9355& 0.9432 \\ \cline{2-9}
       \Xhline{3\arrayrulewidth} 
       \multirow{5}{*}{\textbf{(6F, 2L)}} & {Simple Angle} & 0.6665& 0.6470& 0.6475& 0.6412& 0.6465& 0.6165& 0.7365 \\ \cline{2-9} 
       & {$\mathbf{\frac{\pi}{4}}$ \textbf{angle}} & 0.9798& 0.9815& 0.9810& 0.9815& 0.9810& 0.9895& \textbf{0.9902}\\ \cline{2-9} 
        & {{Entangled Angle}} & 0.8788& 0.8760& 0.8740& 0.8760& 0.8738& 0.8655& 0.8856 \\ \cline{2-9}
        & \textbf{{Amplitude}} & 0.9683& 0.9728& 0.9728& 0.9730& 0.9720& 0.9775& \textbf{0.9793} \\ \cline{2-9}
        & {IQP} & 0.8712& 0.8765& 0.8770& 0.8790& 0.8752& 0.7815& 0.8306\\ \cline{2-9}
       \Xhline{3\arrayrulewidth} 
    \end{tabular}}
    \caption{\textit{The five encodings on the \textbf{MNIST 01 dataset}: training accuracies after 5 epochs, test and F1-Score accuracies for $N$ features and $M$ layers, with $N=\{4,6\}, \  M \in \big\{2, 4\big\}$}.}
\label{tab:MNIST 01_1}
\end{table}

Entangled Angle encoding is better with 4 layers than with 2 layers, whatever the number of features, and on both MNIST datasets. However, on MNIST 08, the best accuracies with this encoding are achieved when $N>4$. The highest accuracies (train accuracy $> 95\%$, test accuracy $96.77\%$ and F1-score $96.71\%$) are achieved with 6 features and 4 layers. Whereas on MNIST01, 4 features were enough to achieve the best accuracies ($> 99\%$). In fact, when compared with the other 4 encodings, Entangled Angle encoding performs better than the others with 6 features and 4 layers on MNIST08. It is important to remember that (6F, 4L) is a case that has given good results with this encoding on previous datasets. And on MNIST01, it is only on (4F,4L) model that Entangled Angle encoding gives the best results. We also note that with this encoding, on all models, train accuracies quickly reach a maximum value (as early as the first two epochs), and stagnate around this value. This is the case for both MNIST datasets. The evolution of these training accuracies can be visualized more easily by viewing \autoref{fig:MNIST01} and \autoref{fig:MNIST08} in the appendix. Of the three angle encodings, Simple Angle encoding gives the best accuracies on cases (4F, 4L) and (8F, 4L). On MNIST08, Entangled Angle encoding is better than the others with 4 layers and for $N\geq 6$. While on MNIST01, it performs better on all cases with 4 layers. In the other cases, $\frac{\pi}{4}$-Angle encoding returns the best precisions.

\begin{table}[H]
    \centering
    \resizebox{\columnwidth}{!}{
    \begin{tabular}{|c|c|c|c|c|c|c|c|c|} 
       \hline
\textbf{Models} & \textbf{Encodings} & \textbf{ep1} & \textbf{ep2} & \textbf{ep3} & \textbf{ep4} & \textbf{ep5} & \textbf{test} & \textbf{F1-score}\\
       \hline
\multirow{5}{*}{\textbf{(8F, 4L)}} & {Simple Angle} & 0.9670& 0.9708& 0.9722& 0.9720& 0.9725& 0.979& 0.9807 \\ \cline{2-9} 
       & { $\frac{\pi}{4}$ {angle}} & 0.9665& 0.9735& 0.9742& 0.9745& 0.9738& 0.972& 0.9743\\ \cline{2-9} 
        & {Entangled Angle} & 0.9670& 0.9860 & 0.9850& 0.9855 & 0.9852 & 0.986 & 0.9870 \\ \cline{2-9}
        & \textbf{{Amplitude}} & 0.9878& 0.9902& 0.9910& 0.9912& 0.9908& 0.9915& \textbf{0.9921} \\ \cline{2-9}
        & {IQP} & 0.9433& 0.9800& 0.9792& 0.9820& 0.9788& 0.9725& 0.9748\\ \cline{2-9}
       \Xhline{3\arrayrulewidth}
       \multirow{5}{*}{\textbf{(8F, 2L)}} & {Simple Angle} & 0.8147& 0.8240& 0.8323& 0.8300& 0.8363& 0.802 & 0.8440 \\ \cline{2-9} 
       & {$\mathbf{\frac{\pi}{4}}$ \textbf{angle}} & 0.9820& 0.9830& 0.9835& 0.9840 & 0.9835& 0.99& \textbf{0.9907} \\ \cline{2-9} 
        & \textbf{Entangled Angle} & 0.9720& 0.9855& 0.9822& 0.9840& 0.9825& 0.971& 0.9736 \\ \cline{2-9}
        & {{Amplitude}} & 0.9517& 0.9550& 0.9540& 0.9547& 0.9545& 0.9465& 0.9524 \\ \cline{2-9}
        & {IQP} & 0.7338& 0.8522& 0.9125& 0.9117& 0.9117& 0.9225& 0.9284\\ \cline{2-9}
       \Xhline{3\arrayrulewidth}
\end{tabular}}
    \caption{\textit{The five encodings on the \textbf{MNIST 01 dataset}: training accuracies after 5 epochs, test and F1-Score accuracies for $N$ features and $M$ layers, with $N=8, \  M \in \big\{2, 4\big\}$}.}
\label{tab:MNIST 01_2}
\end{table}

Amplitude encoding performs best on MNIST08, whatever the model. On MNIST01, this better performance over the other models is only observed on 3 models: (4F, 4L), (6F, 4L) and (8F, 4F). By best performance, we mean a balance between train and test accuracies, followed by a high F1-score (the highest). Given that classification on MNIST08 is more difficult than on MNIST01, the fact that Amplitude encoding is better than the others on MNIST08 could mean that the more complex the problem, the more suitable it is. In fact, on the Malware data, it was one of the best performers.

\renewcommand{\arraystretch}{2.2} %donne la distance entre les lignes%

\begin{table}[H]
    \centering
    \resizebox{\columnwidth}{!}{
    \begin{tabular}{|c|c|c|c|c|c|c|c|c|} 
       \hline
       \small 
\textbf{Models} & \textbf{Encodings} & \textbf{ep1} & \textbf{ep2} & \textbf{ep3} & \textbf{ep4} & \textbf{ep5} & \textbf{test} & \textbf{F1-score}\\
       \hline
        \multirow{5}{*}{\textbf{(4F, 4L)}} & \textbf{Simple Angle} & 0.9525& 0.9530& 0.9535& \textbf{0.9523}& 0.9530 & 0.9508 & 0.9490 \\ \cline{2-9} 
       & {$\frac{\pi}{4}$ {angle}} & 0.8915& 0.8998& 0.8982& 0.8990& 0.8975& 0.9324 & 0.9347 \\ \cline{2-9} 
        & {Entangled Angle} & 0.8775& 0.8890& 0.8878& 0.8868& 0.8878& 0.9252 & 0.9236 \\ \cline{2-9}
        & {{Amplitude}} & 0.9417& 0.9495& 0.9445& 0.9465 & 0.9425 & 0.9350 & 0.9314 \\ \cline{2-9}
        & {\textbf{IQP}} & 0.8968& 0.9055& 0.9048& 0.9060& 0.9030& 0.9426 & \textbf{0.9414} \\ \cline{2-9}
       \Xhline{3\arrayrulewidth}       
       \multirow{5}{*}{\textbf{(4F, 2L)}} & {Simple Angle} & 0.8808 & 0.8942 & 0.8912 & 0.8955 & 0.8960 & 0.9022 & 0.8964 \\ \cline{2-9} 
       & { $\frac{\pi}{4}$-Angle} & 0.8925 & 0.9005 & 0.8995 & 0.9015 & 0.9000 & 0.9309 & 0.9332 \\ \cline{2-9} 
        & {{Entangled Angle}} & 0.8762 & 0.8888 & 0.8865 & 0.8888 & 0.8900 & 0.9191 & 0.9171 \\ \cline{2-9}
        & {\textbf{Amplitude}} & 0.9433 & 0.9417 & 0.9453 & 0.9460 & 0.9447 & 0.9411 & \textbf{0.9483} \\ \cline{2-9}
        & {IQP} & 0.8990 & 0.9052 & 0.9025 & 0.9093 & 0.9058 & 0.9442 & 0.9429 \\ \cline{2-9}
       \Xhline{3\arrayrulewidth}    
       \multirow{5}{*}{\textbf{(6F, 4L)}} & {Simple Angle} & 0.9320& 0.9395& 0.9443& 0.9437& 0.9453& 0.9498& 0.9496 \\ \cline{2-9} 
       & { $\frac{\pi}{4}$-Angle} & 0.9237& 0.9185& 0.9273& 0.9447& 0.9467& 0.9252& 0.9259\\ \cline{2-9} 
        & {\textbf{Entangled Angle}} & 0.9450& 0.9543& 0.9565& 0.9553& 0.9563& 0.9677& \textbf{0.9671} \\ \cline{2-9}
        & {\textbf{Amplitude}} & 0.9580& 0.9543& 0.9527& 0.9567& 0.9567& 0.9570& \textbf{0.9560} \\ \cline{2-9}
        & {IQP} & 0.9283& 0.9353& 0.9345& 0.9350& 0.9355& 0.9447& 0.9457 \\ \cline{2-9}
       \Xhline{3\arrayrulewidth} 
       \multirow{5}{*}{\textbf{(6F, 2L)}} & \textbf{Simple Angle} & 0.9123& 0.9350& 0.9363& 0.9363& 0.9355& 0.9401& \textbf{0.9399} \\ \cline{2-9} 
       & { $\frac{\pi}{4}$-Angle} & 0.9062& 0.9133& 0.9123& 0.9145& 0.9145& 0.9488& 0.9489 \\ \cline{2-9} 
        & {{Entangled Angle}} & 0.8380& 0.8390& 0.8405& 0.8385& 0.8397& 0.8520& 0.8688 \\ \cline{2-9}
        & {\textbf{Amplitude}} & 0.9397& 0.9527& 0.9525& 0.9513& 0.9515& 0.9401& \textbf{0.9369} \\ \cline{2-9}
        & {IQP} & 0.8718& 0.8618& 0.8605& 0.8612& 0.8645& 0.8024& 0.7535\\ \cline{2-9}
       \Xhline{3\arrayrulewidth} 
    \end{tabular}}
    \caption{\textit{The five encodings on the \textbf{MNIST 08 dataset}: training accuracies after 5 epochs, test and F1-Score accuracies for $N$ features and $M$ layers, with $N=\{4,6\}, \  M \in \big\{2, 4\big\}$}.}
\label{tab:MNIST 08_1}
\end{table}

\begin{table}[H]
    \centering
    \resizebox{\columnwidth}{!}{
    \begin{tabular}{|c|c|c|c|c|c|c|c|c|} 
       \hline
\textbf{Models} & \textbf{Encodings} & \textbf{ep1} & \textbf{ep2} & \textbf{ep3} & \textbf{ep4} & \textbf{ep5} & \textbf{test} & \textbf{F1-score}\\
       \hline
\multirow{5}{*}{\textbf{(8F, 4L)}} & {Simple Angle} & 0.9390& 0.9323& 0.9303& 0.9320& 0.9325 & 0.9406& 0.9415 \\ \cline{2-9} 
       & { $\frac{\pi}{4}$ {angle}} & 0.8898& 0.9233& 0.9147& 0.9287& 0.9243& 0.9211& 0.9215 \\ \cline{2-9} 
        & {Entangled Angle} & 0.9295& 0.9473& 0.9490& 0.9517& 0.9525& 0.9580& 0.9577 \\ \cline{2-9}
        & {\textbf{Amplitude}} & 0.9610& 0.9657& 0.9680& 0.9690& 0.9670& 0.9641& \textbf{0.9638} \\ \cline{2-9}
        & {IQP} & 0.8620& 0.9010& 0.9183& 0.9247& 0.9270& 0.9365& 0.9381 \\ \cline{2-9}
       \Xhline{3\arrayrulewidth}
       \multirow{5}{*}{\textbf{(8F, 2L)}} & {Simple Angle} & 0.8675& 0.9327& 0.9320& 0.9325& 0.9313& 0.9385& 0.9385 \\ \cline{2-9} 
       & {$\frac{\pi}{4}$ {angle}} & 0.8988& 0.9070& 0.9093& 0.9105& 0.9067& 0.9273& 0.9294 \\ \cline{2-9} 
        & {Entangled Angle} & 0.8588& 0.8835& 0.8788& 0.8785& 0.8822& 0.9058& 0.9090\\ \cline{2-9}
        & {\textbf{Amplitude}} & 0.9367& 0.9373& 0.9373& 0.9363& 0.9380& 0.9431& \textbf{0.9414} \\ \cline{2-9}
        & {IQP} & 0.8165& 0.8297& 0.8295& 0.8237& 0.8230& 0.8776& 0.8777\\ \cline{2-9}
       \Xhline{3\arrayrulewidth}
\end{tabular}}
    \caption{\textit{The five encodings on the \textbf{MNIST 08 dataset}: training accuracies after 5 epochs, test and F1-Score accuracies for $N$ features and $M$ layers, with $N=8, \  M \in \big\{2, 4\big\}$}.}
\label{tab:MNIST 08_2}
\end{table}

\section{Conclusion and perspectives}\label{sec:conclusion}
The aim of our work was to contribute to the optimisation of QML algorithms in the current era of NISQ. Indeed, knowing that in QML, the data encoding part plays a remarkable role on the performance of the model \cite{Schuld_20211}\cite{Caro_20211}, the choice of a suitable encoding method is important for the construction of a high-performance model. In order to make a significant contribution to this choice of encoding method, we have selected five types of encoding that are fairly recurrent in the literature \cite{math12213318}\cite{rath2023quantumdataencodingcomparative}\cite{munikote2024comparingquantumencodingtechniques}. Starting from a Quantum Neural Networks (QNN) model (\autoref{fig:single model}), we benchmarked these encodings on 4 datasets. 

The results obtained show that the performance of the encoding methods varies significantly depending on the dataset used and the specific features of the models, in particular the number of features and layers.  In general, the results show that, encodings such as $\frac{\pi}{4}$-Angle  and IQP encodings need few features to return good accuracies. To improve these accuracies, increasing the number of layers is more of a solution than increasing the number of features. 

Simple Angle encoding quickly achieves maximum training accuracy from the very first epochs. This accuracy will either remain constant over the following epochs, or decrease at some point and return to the maximum epoch achieved at the beginning. It is therefore an encoding that does not require several epochs for training. However, of the three angle encodings, it is the one that performs the worst on the average of the models. The performance of these angle encodings can be seen in the WDBC dataset, where Entangled Angle encoding was one of the best-performing encodings on this dataset. 
On the other hand, the Amplitude encoding was better on several models on the Malware and MNIST datasets, while on WDBC, it returned the lowest performance. These performances tend to decrease as the number of features decreases. This suggests that not only does this encoding need a large amount of data to return good results, but in a case with fewer data, the number of features must be increased. 

In summary for datasets with features ranked in order of importance, if the size of the dataset is large enough (thousands of data points), we can build a good model based on Amplitude encoding with very few features, and therefore very few qubits. However, to ensure good accuracy, it would be preferable to choose a number of features as a power of 2. In a case where the dataset does not have enough data (as in the case of the WDBC dataset), Angle encodings will be more suitable for a high-performance model, also using fewer qubits. $\frac{\pi}{4}$-Angle and Entangled Angle encoding returned the best results compared with the Simple Angle encoding, while the latter requires little epoch to achieve its best accuracies. However, in the three cases,  it is important to choose a large enough number of layers (at least 4 layers) to have good performances.

It should be remembered that we have used a classification model with $\RY$ rotation gates followed by CNOT gates (\autoref{fig:single model}). It is therefore possible that some encodings, such as Amplitude and $\frac{\pi}{4}$ Angle encodings have architectures that fit well with this classification model. Another study could consist of choosing these encodings that gave us the best results, and benchmarking them on more than one QNN model (using different quantum gates). In this way, we could observe how these encodings behave with other QNN architectures.

\printbibliography

@ARTICLE{2021QuIP...20..119C,
       author = {{Chalumuri}, Avinash and {Kune}, Raghavendra and {Manoj}, B.~S.},
        title = "{A hybrid classical-quantum approach for multi-class classification}",
      journal = {Quantum Information Processing},
         year = 2021,
        month = mar,
       volume = {20},
       number = {3},
          eid = {119},
        pages = {119},
          doi = {10.1007/s11128-021-03029-9},
       adsurl = {https://ui.adsabs.harvard.edu/abs/2021QuIP...20..119C}
}

@article{Preskill_2018,
   title={Quantum {C}omputing in the {NISQ} era and beyond},
   volume={2},
   ISSN={2521-327X},
   url={http://dx.doi.org/10.22331/q-2018-08-06-79},
   DOI={10.22331/q-2018-08-06-79},
   journal={Quantum},
   publisher={Verein zur Forderung des Open Access Publizierens in den Quantenwissenschaften},
   author={Preskill, John},
   year={2018},
   month=aug, pages={79} }

@misc{breast,

    doi  = {https://doi.org/10.24432/C5DW2B},

    year  = 1993,

    publisher = {UCI Machine Learning Repository},

    author = {William Wolberg and Olvi Mangasarian and Nick Street},

    title  = {Breast {C}ancer {W}isconsin ({Diagnostic})},
}

@article{PhysRevLett,
  title = {Quantum {M}achine {L}earning in {F}eature {H}ilbert {S}paces},
  author = {Schuld, Maria and Killoran, Nathan},
  journal = {Phys. Rev. Lett.},
  volume = {122},
  issue = {4},
  pages = {040504},
  numpages = {6},
  year = {2019},
  month = {2},
  publisher = {American Physical Society},
  doi = {10.1103/PhysRevLett.122.040504},
  url = {https://link.aps.org/doi/10.1103/PhysRevLett.122.040504}
}

@Article{app9153169,
AUTHOR = {Baldominos, Alejandro and Saez, Yago and Isasi, Pedro},
TITLE = {A {S}urvey of {H}andwritten {C}haracter {R}ecognition with {MNIST} and {EMNIST}},
JOURNAL = {Applied Sciences},
VOLUME = {9},
YEAR = {2019},
NUMBER = {15},
ARTICLE-NUMBER = {3169},
URL = {https://www.mdpi.com/2076-3417/9/15/3169},
ISSN = {2076-3417},
}

@article{Skolik_2021,
   title={Layerwise learning for quantum neural networks},
   volume={3},
   ISSN={2524-4914},
   url={http://dx.doi.org/10.1007/s42484-020-00036-4},
   DOI={10.1007/s42484-020-00036-4},
   number={1},
   journal={Quantum Machine Intelligence},
   publisher={Springer Science and Business Media LLC},
   author={Skolik Andrea and McClean Jarrod R. and Mohseni Masoud and van der Smagt Patrick and Leib Martin},
   year={2021},
   month={1},
}

@misc{deluca2024empiricalpowerquantumencoding,
      title={Empirical {P}ower of {Q}uantum {E}ncoding {M}ethods for {B}inary {C}lassification}, 
      author={Gennaro De Luca and Andrew Vlasic and Michael Vitz and Anh Pham},
      year={2024},
      eprint={2408.13109},
      archivePrefix={arXiv},
      primaryClass={quant-ph},
      url={https://arxiv.org/abs/2408.13109}, 
}

@article{Havl_ek_2019,
   title={Supervised learning with quantum-enhanced feature spaces},
   volume={567},
   ISSN={1476-4687},
   url={http://dx.doi.org/10.1038/s41586-019-0980-2},
   DOI={10.1038/s41586-019-0980-2},
   number={7747},
   journal={Nature},
   publisher={Springer Science and Business Media LLC},
   author={Havlicek, Vojtěch and Corcoles, Antonio D. and Temme, Kristan and Harrow, Aram W. and Kandala, Abhinav and Chow, Jerry M. and Gambetta, Jay M.},
   year={2019},
   month=mar, pages={209–212} }

@article{Bremner_2010,
   title={Classical simulation of commuting quantum computations implies collapse of the polynomial hierarchy},
   volume={467},
   ISSN={1471-2946},
   url={http://dx.doi.org/10.1098/rspa.2010.0301},
   DOI={10.1098/rspa.2010.0301},
   number={2126},
   journal={Proceedings of the Royal Society A: Mathematical, Physical and Engineering Sciences},
   publisher={The Royal Society},
   author={Bremner, Michael J. and Jozsa, Richard and Shepherd, Dan J.},
   year={2010},
   month=aug, pages={459–472} }

@article{articleHong,
author = {Hong, Ying-Yi and Arce, Christine and Huang, Tsung-Wei},
year = {2023},
month = {01},
pages = {1-1},
title = {A {R}obust {H}ybrid {C}lassical and {Q}uantum {M}odel for {S}hort-{T}erm {W}ind {S}peed {F}orecasting},
volume = {PP},
journal = {IEEE Access},
doi = {10.1109/ACCESS.2023.3308053}
}

@article{araujo,
author = {Araujo, Israel and Park, Daniel and Petruccione, Francesco and da Silva, Adenilton},
year = {2020},
month = {08},
pages = {},
title = {A divide-and-conquer algorithm for quantum state preparation},
doi = {10.48550/arXiv.2008.01511}
}

@article{Schuld_20211,
   title={Effect of data encoding on the expressive power of variational quantum-machine-learning models},
   volume={103},
   ISSN={2469-9934},
   url={http://dx.doi.org/10.1103/PhysRevA.103.032430},
   DOI={10.1103/physreva.103.032430},
   number={3},
   journal={Physical Review A},
   publisher={American Physical Society (APS)},
   author={Schuld, Maria and Sweke, Ryan and Meyer, Johannes Jakob},
   year={2021},
   month={3},
}

@article{Caro_20211,
   title={Encoding-dependent generalization bounds for parametrized quantum circuits},
   volume={5},
   ISSN={2521-327X},
   url={http://dx.doi.org/10.22331/q-2021-11-17-582},
   DOI={10.22331/q-2021-11-17-582},
   journal={Quantum},
   publisher={Verein zur Forderung des Open Access Publizierens in den Quantenwissenschaften},
   author={Caro, Matthias C. and Gil-Fuster, Elies and Meyer, Johannes Jakob and Eisert, Jens and Sweke, Ryan},
   year={2021},
   month={11},
pages={582}}

@article{Low_2024,
   title={Trading {T} gates for dirty qubits in state preparation and unitary synthesis},
   volume={8},
   ISSN={2521-327X},
   url={http://dx.doi.org/10.22331/q-2024-06-17-1375},
   DOI={10.22331/q-2024-06-17-1375},
   journal={Quantum},
   publisher={Verein zur Forderung des Open Access Publizierens in den Quantenwissenschaften},
   author={Low, Guang Hao and Kliuchnikov, Vadym and Schaeffer, Luke},
   year={2024},
   month=jun, pages={1375} }

@misc{kerenidis2016quantumrecommendationsystems,
      title={Quantum {R}ecommendation {S}ystems}, 
      author={Iordanis Kerenidis and Anupam Prakash},
      year={2016},
      eprint={1603.08675},
      archivePrefix={arXiv},
      primaryClass={quant-ph},
      url={https://arxiv.org/abs/1603.08675}, 
}

@misc{autoqml,
      title={AutoQML: {A}utomated {Q}uantum {M}achine {L}earning for {W}i-{F}i {I}ntegrated {S}ensing and {C}ommunications}, 
      author={Toshiaki Koike-Akino and Pu Wang and Ye Wang},
      year={2022},
      eprint={2205.09115},
      archivePrefix={arXiv},
      primaryClass={cs.LG},
      url={https://arxiv.org/abs/2205.09115}, 
}

@article{Alvarez_Rodriguez_2017,
   title={Supervised {Q}uantum {L}earning without {M}easurements},
   volume={7},
   ISSN={2045-2322},
   url={http://dx.doi.org/10.1038/s41598-017-13378-0},
   DOI={10.1038/s41598-017-13378-0},
   number={1},
   journal={Scientific Reports},
   publisher={Springer Science and Business Media LLC},
   author={Alvarez-Rodriguez, Unai and Lamata, Lucas and Escandell-Montero, Pablo and Martín-Guerrero, José D. and Solano, Enrique},
   year={2017},
   month=oct }

@book{Nielsen_Chuang_2010, place={Cambridge}, title={{Q}uantum {C}omputation and {Q}uantum {I}nformation: 10th {A}nniversary {E}dition}, publisher={Cambridge University Press}, author={Nielsen, Michael A. and Chuang, Isaac L.}, year={2010}}

@article{jolliffe,
author = {Jolliffe, Ian and Cadima, Jorge},
year = {2016},
month = {04},
pages = {20150202},
title = {Principal component analysis: {A} review and recent developments},
volume = {374},
journal = {Philosophical Transactions of the Royal Society A: Mathematical, Physical and Engineering Sciences},
doi = {10.1098/rsta.2015.0202}
}

@article{Perez_Salinas_2020,
   title={Data re-uploading for a universal quantum classifier},
   volume={4},
   ISSN={2521-327X},
   url={http://dx.doi.org/10.22331/q-2020-02-06-226},
   DOI={10.22331/q-2020-02-06-226},
   journal={Quantum},
   publisher={Verein zur Forderung des Open Access Publizierens in den Quantenwissenschaften},
   author={Pérez-Salinas, Adrián and Cervera-Lierta, Alba and Gil-Fuster, Elies and Latorre, José I.},
   year={2020},
   month=feb, pages={226} }

@misc{quantumcomputingqiskit,
      title={Quantum computing with {Q}iskit}, 
      author={Ali Javadi-Abhari and Matthew Treinish and Kevin Krsulich and Christopher J. Wood and Jake Lishman and Julien Gacon and Simon Martiel and Paul D. Nation and Lev S. Bishop and Andrew W. Cross and Blake R. Johnson and Jay M. Gambetta},
      year={2024},
      eprint={2405.08810},
      archivePrefix={arXiv},
      primaryClass={quant-ph},
      url={https://arxiv.org/abs/2405.08810}, 
}

@Article{math12213318,
AUTHOR = {Ranga, Deepak and Rana, Aryan and Prajapat, Sunil and Kumar, Pankaj and Kumar, Kranti and Vasilakos, Athanasios V.},
TITLE = {Quantum {M}achine {L}earning: {E}xploring the {R}ole of {D}ata {E}ncoding {T}echniques, {C}hallenges, and {F}uture {D}irections},
JOURNAL = {Mathematics},
VOLUME = {12},
YEAR = {2024},
NUMBER = {21},
ARTICLE-NUMBER = {3318},
URL = {https://www.mdpi.com/2227-7390/12/21/3318},
ISSN = {2227-7390},
DOI = {10.3390/math12213318}
}

@misc{rath2023quantumdataencodingcomparative,
      title={Quantum {D}ata {E}ncoding: {A} {C}omparative {A}nalysis of {C}lassical-to-{Q}uantum {M}apping {T}echniques and {T}heir {I}mpact on {M}achine {L}earning {A}ccuracy}, 
      author={Minati Rath and Hema Date},
      year={2023},
      eprint={2311.10375},
      archivePrefix={arXiv},
      primaryClass={quant-ph},
      url={https://arxiv.org/abs/2311.10375}, 
}

@misc{munikote2024comparingquantumencodingtechniques,
      title={Comparing {Q}uantum {E}ncoding {T}echniques}, 
      author={Nidhi Munikote},
      year={2024},
      eprint={2410.09121},
      archivePrefix={arXiv},
      primaryClass={quant-ph},
      url={https://arxiv.org/abs/2410.09121}, 
}

@Article{e25020287,
AUTHOR = {Zeguendry, Amine and Jarir, Zahi and Quafafou, Mohamed},
TITLE = {Quantum {M}achine {L}earning: {A} {R}eview and {C}ase {S}tudies},
JOURNAL = {Entropy},
VOLUME = {25},
YEAR = {2023},
NUMBER = {2},
ARTICLE-NUMBER = {287},
URL = {https://www.mdpi.com/1099-4300/25/2/287},
PubMedID = {36832654},
ISSN = {1099-4300},
DOI = {10.3390/e25020287}
}

@misc{heese2025explainingquantumcircuitsshapley,
      title={Explaining {Q}uantum {C}ircuits with {S}hapley {V}alues: {T}owards {E}xplainable {Q}uantum {M}achine {L}earning}, 
      author={Raoul Heese and Thore Gerlach and Sascha Mücke and Sabine Müller and Matthias Jakobs and Nico Piatkowski},
      year={2025},
      eprint={2301.09138},
      archivePrefix={arXiv},
      primaryClass={quant-ph},
      url={https://arxiv.org/abs/2301.09138}, 
}

@INPROCEEDINGS{9474321,
  author={Yang, Limin and Ciptadi, Arridhana and Laziuk, Ihar and Ahmadzadeh, Ali and Wang, Gang},
  booktitle={2021 IEEE Security and Privacy Workshops (SPW)}, 
  title={{BODMAS}: {A}n {O}pen {D}ataset for {L}earning based {T}emporal {A}nalysis of {PE} {M}alware}, 
  year={2021},
  volume={},
  number={},
  pages={78-84},
  doi={10.1109/SPW53761.2021.00020}}

@misc{PEML,
      title={{PE} {M}alware {M}achine {L}earning {D}ataset}, 
      author={Michael Lester},
      url={https://practicalsecurityanalytics.com/pe-malware-machine-learning-dataset/}, 
}

@phdthesis{marais:tel-04416984,
  TITLE = {{Am{\'e}liorations des outils de d{\'e}tection de malwares par analyse statique gr{\^a}ce {\`a} des m{\'e}canismes d'intelligence artificielle}},
  AUTHOR = {Marais, Benjamin},
  URL = {https://theses.hal.science/tel-04416984},
  NUMBER = {2023NORMC245},
  SCHOOL = {{Normandie Universit{\'e}}},
  YEAR = {2023},
  MONTH = Dec,
  TYPE = {Theses},
  HAL_ID = {tel-04416984},
  HAL_VERSION = {v1},
}

@misc{bermejo2024quantumconvolutionalneuralnetworks,
      title={Quantum Convolutional Neural Networks are (Effectively) Classically Simulable}, 
      author={Pablo Bermejo and Paolo Braccia and Manuel S. Rudolph and Zoë Holmes and Lukasz Cincio and M. Cerezo},
      year={2024},
      eprint={2408.12739},
      archivePrefix={arXiv},
      primaryClass={quant-ph},
      url={https://arxiv.org/abs/2408.12739}, 
}

@article{Shepherd_2009,
   title={Temporally unstructured quantum computation},
   volume={465},
   ISSN={1471-2946},
   url={http://dx.doi.org/10.1098/rspa.2008.0443},
   DOI={10.1098/rspa.2008.0443},
   number={2105},
   journal={Proceedings of the Royal Society A: Mathematical, Physical and Engineering Sciences},
   publisher={The Royal Society},
   author={Shepherd, Dan and Bremner, Michael J.},
   year={2009},
   month=feb, pages={1413–1439} }

\section{Appendix}\label{sec:appendix}
In this appendix, we present a benchmark of single-angle encoding used with RX, then with RY. This benchmarking allows us to evaluate how Simple Angle encoding behaves, depending on the rotation gate used.

We also present graphical visualisations of the training results obtained during the evaluation of five encoding methods. Although the detailed results were presented in tabular form in the main article, the following figures illustrate the evolution of performance, allowing trends and variations over time to be better appreciated.

\normalsize
\subsection*{Benchmarking Simple Angle encoding with $\RY$ and $\RX$}
\begin{figure}[H]
    \centering
    \includegraphics[width=0.92\linewidth]{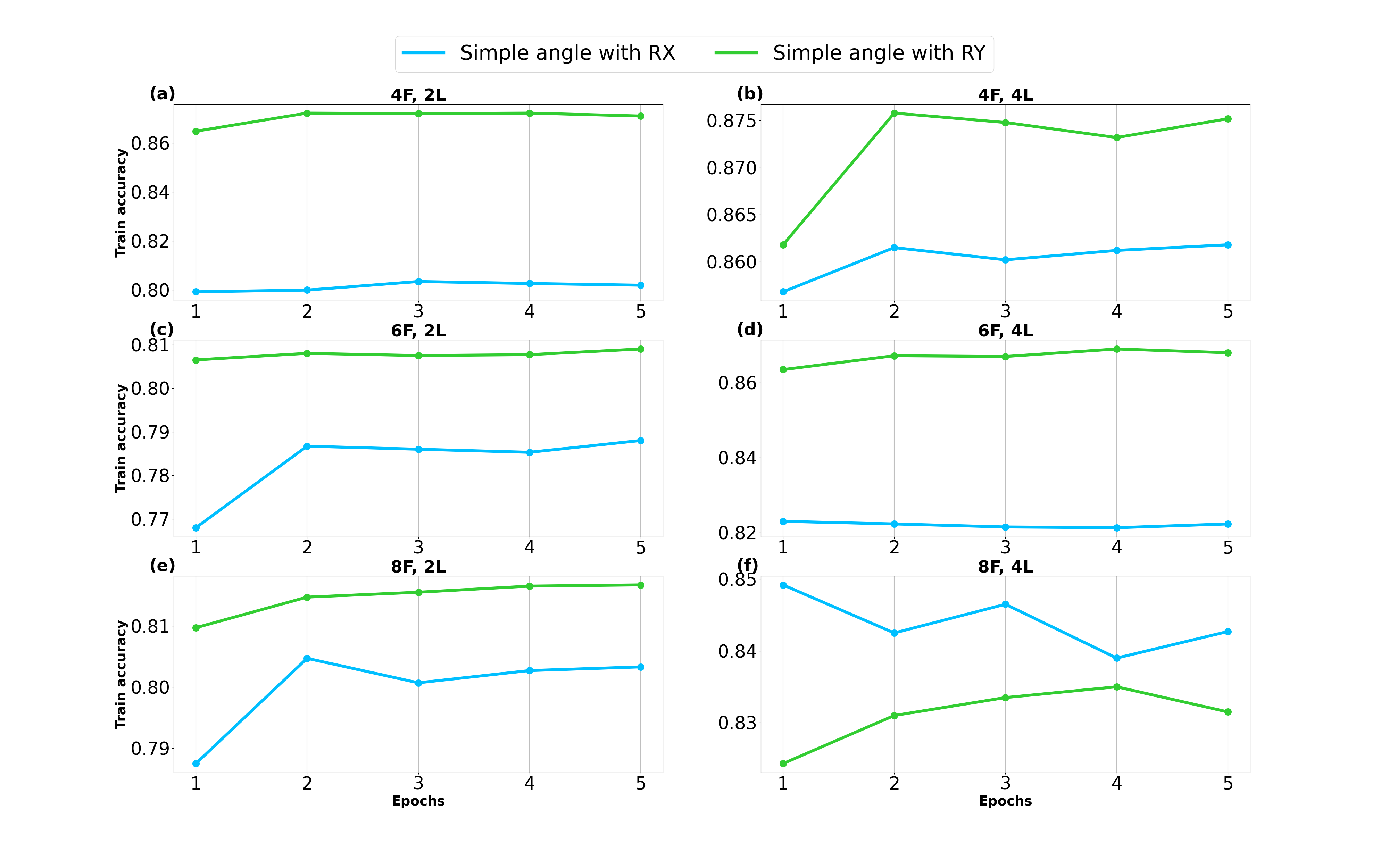}
    \caption{\textit{Train accuracies after 5 epochs: benchmarking Simple Angle encoding with $\RX$ and $\RY$ on  \textbf{Malware dataset.}}}
    \label{fig:Malware_RXRY}
\end{figure}

\begin{figure}[H]
    \centering
    \includegraphics[width=0.92\linewidth]{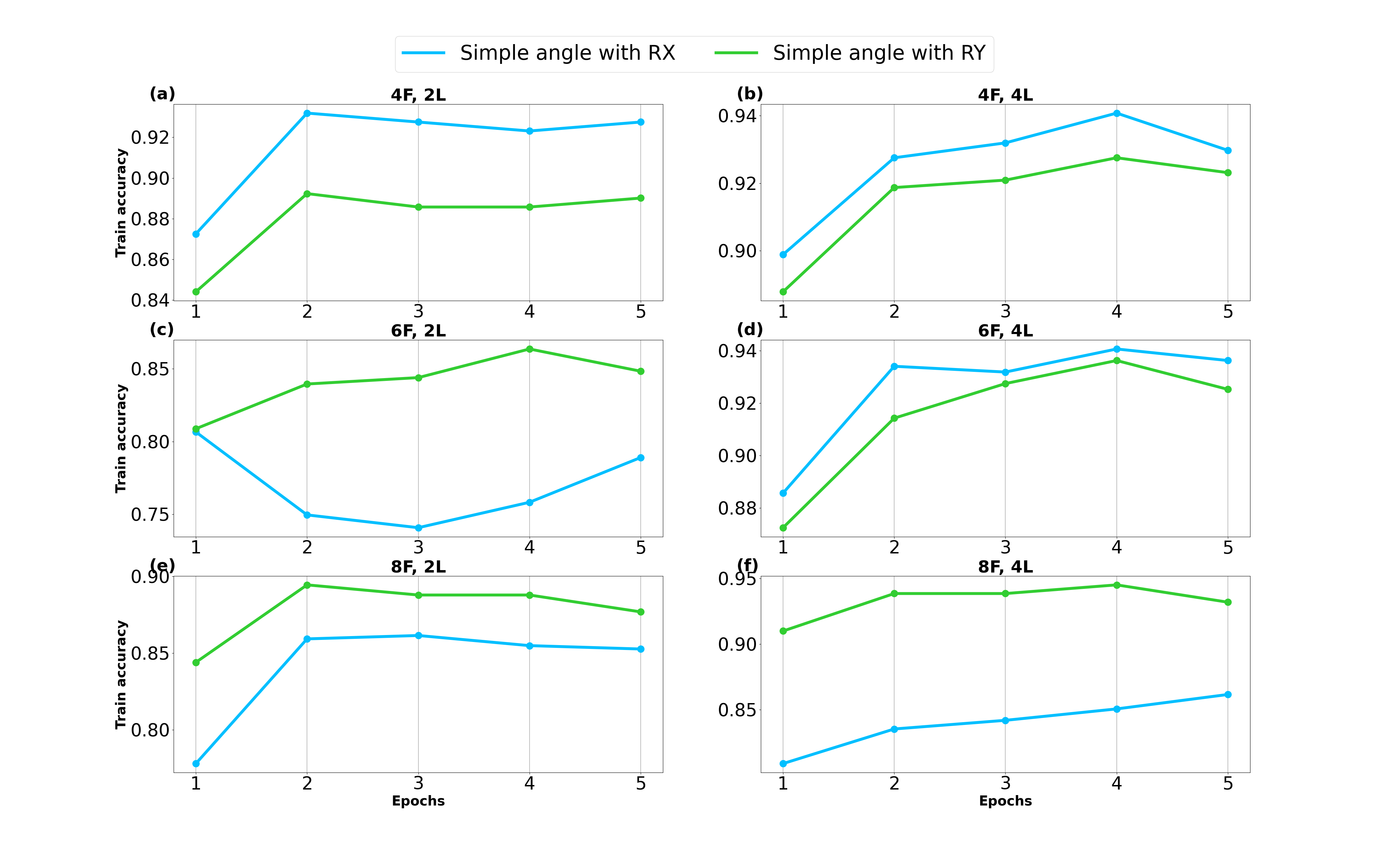}
    \caption{\textit{Train accuracies after 5 epochs: benchmarking Simple Angle encoding with $\RX$ and $\RY$ on  \textbf{WDBC dataset.}}}
    \label{fig:Breast_RXRY}
\end{figure}

\begin{figure}[H]
    \centering
    \includegraphics[width=0.92\linewidth]{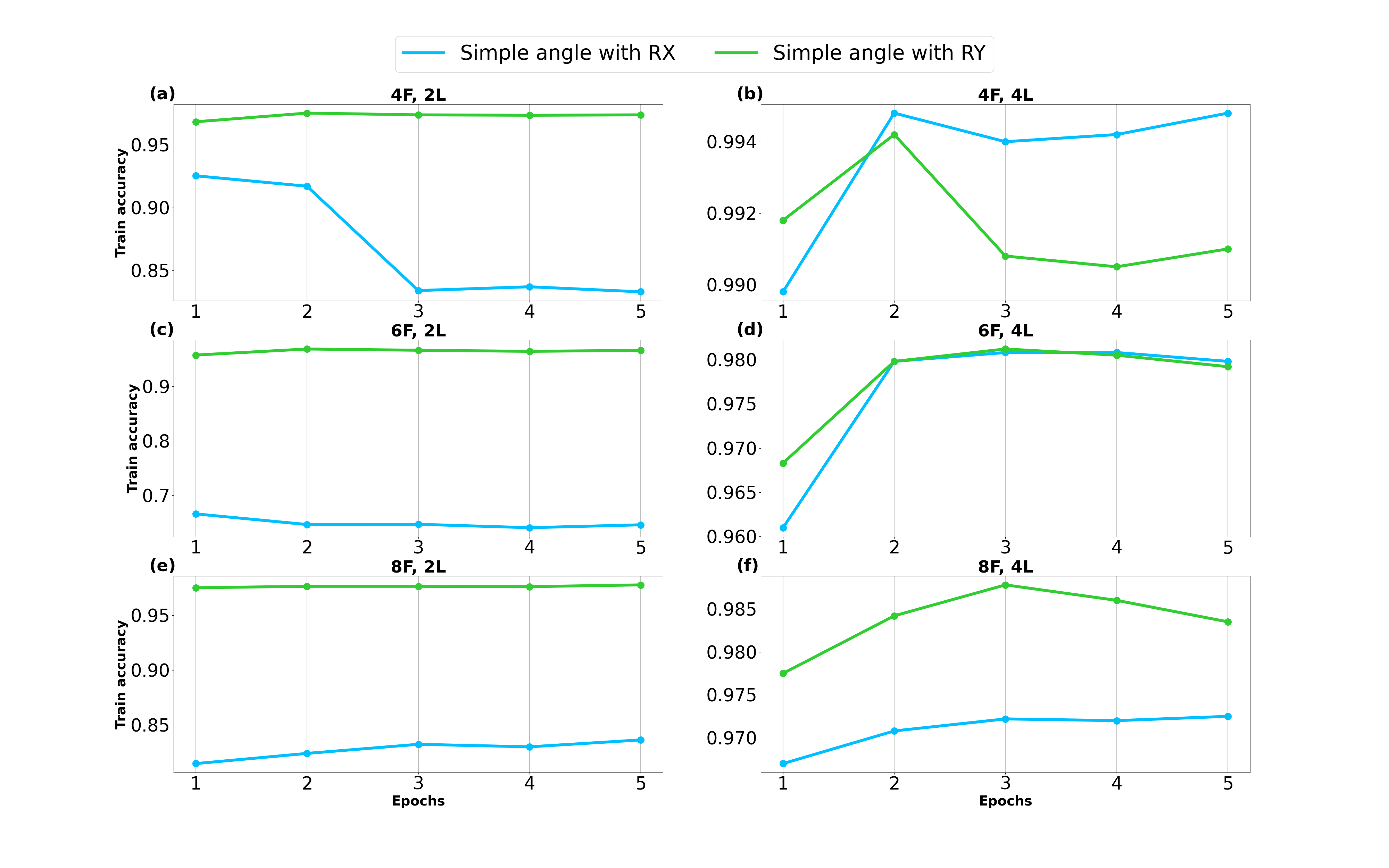}
    \caption{\textit{Train accuracies after 5 epochs: benchmarking Simple Angle encoding with $\RX$ and $\RY$ on  \textbf{MNIST01 dataset.}}}
    \label{fig:MNIST01_RXRY}
\end{figure}

\begin{figure}[H]
    \centering
    \includegraphics[width=0.92\linewidth]{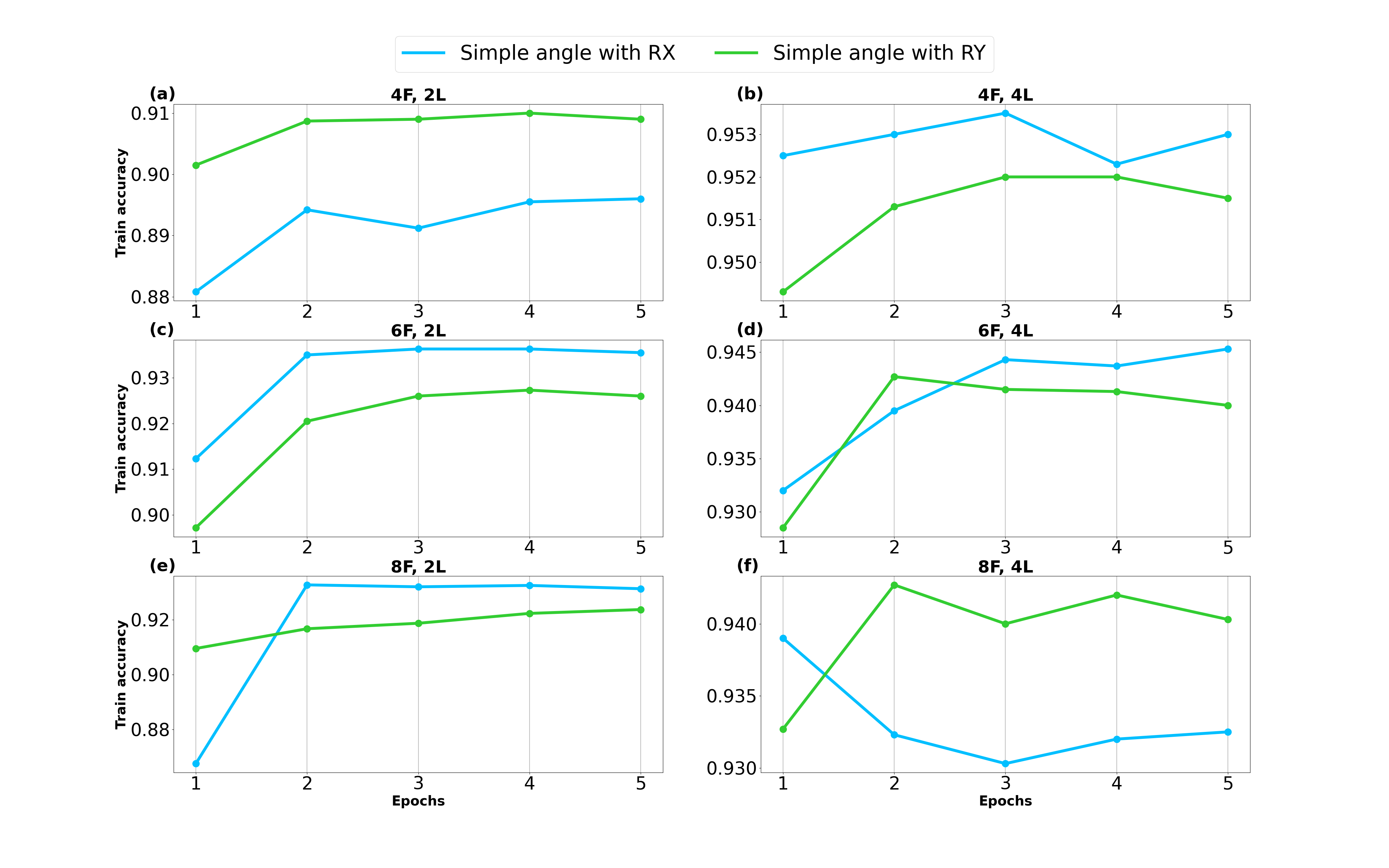}
    \caption{\textit{Train accuracies after 5 epochs: benchmarking Simple Angle encoding with $\RX$ and $\RY$ on  \textbf{MNIST08 dataset.}}}
    \label{fig:MNIST08_RXRY}
\end{figure}

We compare Simple Angle encoding with $\RX$ (results presented in the previous section) and Simple Angle encoding with $\RY$ on the 4 datasets as presented in equations \ref{eq:Simple Angle}. With the simple $\RY$ angle encoding, we performed the same simulations as those done with the five encodings compared in \autoref{sec:results}. In this subsection, we present the results on the 4 datasets in \autoref{tab:malware_WDBC_rx_vs_ry} and \autoref{tab:MNIST01_MNIST08_rx_vs_ry}, and Figures \ref{fig:Breast_RXRY}, \ref{fig:Malware_RXRY}, \ref{fig:MNIST01_RXRY} and \ref{fig:MNIST08_RXRY} present the graphical visualizations.

\begin{table}[H]
    \centering
    \small
    \resizebox{\columnwidth}{!}{
    \begin{tabular}{|c|c|c|c|c|c|c|c|c|c|} 
       \hline
       Dataset & Models & Simple Angle & e1 & e2 & e3 & e4 & e5 & Test & F1-score  \\
       \Xhline{4\arrayrulewidth}
       \multirow{12}{*}{\textbf{Malware}} & \multirow{2}{*}{4F, 4L} & With $\RX$& 0.8568& 0.8615& 0.8602& 0.8612& 0.8618& 0.877& 0.8806 \\ \cline{3-10}
       & & With $\RY$& 0.8618 & 0.8758 & 0.8748 & 0.8732 & 0.8752 & 0.865 & 0.8710 \\ \Xcline{2-10}{3\arrayrulewidth}
      & \multirow{2}{*}{\textbf{4F, 2L}} & With $\RX$& 0.7993& 0.8000& 0.8035& 0.8027& 0.8020& 0.8155& 0.8305 \\ \cline{3-10}
      & & \textbf{With} $\mathbf{\RY}$& 0.8648 & 0.8722 & 0.8720 & 0.8722 &  0.8710 & 0.861  & \textbf{0.8677} \\  \Xcline{2-10}{3\arrayrulewidth}
       &\multirow{2}{*}{\textbf{6F, 4L}} & With $\RX$& 0.8230& 0.8223& 0.8215& 0.8213& 0.8223& 0.8475& 0.8563 \\ \cline{3-10}
      & & \textbf{With} $\mathbf{\RY}$ & 0.8635 & 0.8672 & 0.8670 & 0.8690 & 0.8680 & 0.866 & \textbf{0.8718} \\  \Xcline{2-10}{3\arrayrulewidth}
      & \multirow{2}{*}{6F, 2L} & With $\RX$& 0.7680& 0.7867& 0.7860& 0.7853& 0.7880& 0.811& 0.8288 \\ \cline{3-10}
      & & With $\RY$& 0.8065 & 0.8080 & 0.8075 & 0.8077 & 0.8090 & 0.82 & 0.8357 \\  \Xcline{2-10}{3\arrayrulewidth}
       &\multirow{2}{*}{8F, 4L} & With $\RX$& 0.8492& 0.8425 & 0.8465& 0.8390& 0.8427& 0.84857& 0.8574 \\  \cline{3-10}
      & & With $\RY$& 0.8243 & 0.8310 & 0.8335 & 0.8350 & 0.8315 & 0.8315 & 0.8449 \\ \Xcline{2-10}{3\arrayrulewidth}
      & \multirow{2}{*}{8F, 2L} & With $\RX$& 0.7875& 0.8047& 0.8007& 0.8027& 0.8033 & 0.8195 & 0.8353 \\  \cline{3-10}
     &  & With $\RY$& 0.8097 & 0.8147 & 0.8155 & 0.8165 & 0.8167 & 0.8235 & 0.8379 \\
       \Xhline{4\arrayrulewidth}
       \multirow{12}{*}{\textbf{WDBC}} & \multirow{2}{*}{4F, 4L} & With $\RX$& 0.8989& 0.9275& 0.9319& 0.9407& 0.9297& 0.9473& 0.9230 \\ \cline{3-10}
      & & With $\RY$& 0.8879 & 0.9187 & 0.9209 & 0.9275 & 0.9231 & 0.9473 & 0.9268 \\ \Xcline{2-10}{3\arrayrulewidth}
      & \multirow{2}{*}{4F, 2L} & With $\RX$& 0.8725& 0.9319& 0.9275& 0.9231& 0.9275& 0.9122& 0.8888 \\  \cline{3-10}
      & & With $\RY$& 0.8440 & 0.8923 & 0.8857 & 0.8857 & 0.8901 & 0.9298 & 0.9\\  \Xcline{2-10}{3\arrayrulewidth}
      & \multirow{2}{*}{6F, 4L} & With $\RX$& 0.8857& 0.9341& 0.9319& 0.9407& 0.9363& 0.9300& 0.9135 \\  \cline{3-10}
      & & With $\RY$& 0.8725 & 0.9143 & 0.9275 & 0.9363 & 0.9253 & 0.9385 & 0.9135 \\  \Xcline{2-10}{3\arrayrulewidth}
      & \multirow{2}{*}{6F, 2L} & With $\RX$& 0.8066& 0.7495& 0.7407& 0.7582& 0.7890& 0.8333& 0.7164 \\  \cline{3-10}
      & & With $\RY$& 0.8088 & 0.8396 & 0.8440 & 0.8637 & 0.8484 & 0.8684 & 0.7945 \\ \Xcline{2-10}{3\arrayrulewidth}
      & \multirow{2}{*}{\textbf{8F, 4L}} & With $\RX$& 0.8088& 0.8352& 0.8418& 0.8505& 0.8615& 0.8508& 0.7536 \\  \cline{3-10}
      & & \textbf{With} $\mathbf{\RY}$ & 0.9099 & 0.9385 & 0.9385 & 0.9451 & 0.9319 & 0.9473 & \textbf{0.9285} \\  \Xcline{2-10}{3\arrayrulewidth}
      & \multirow{2}{*}{\textbf{8F, 2L}} & With $\RX$& 0.7780& 0.8593& 0.8615& 0.8549& 0.8527& 0.8596& 0.8048 \\  \cline{3-10}
     &  & \textbf{With} $\mathbf{\RY}$ & 0.8440 & 0.8945 & 0.8879 & 0.8879 & 0.8769 & 0.9122 & \textbf{0.8837} \\
       \Xhline{4\arrayrulewidth}
    \end{tabular}}
    \normalsize
    \caption{\textit{Benchmarking Simple Angle encoding with $\RX$ and $\RY$ on  \textbf{Malware} and \textbf{WDBC datasets}.}}
    \label{tab:malware_WDBC_rx_vs_ry}
\end{table}

\newpage

\begin{table}[H]
    \centering
    \small
    \resizebox{\columnwidth}{!}{
    \begin{tabular}{|c|c|c|c|c|c|c|c|c|c|} 
       \hline
      Dataset& Models & Simple Angle encoding & e1 & e2 & e3 & e4 & e5 & Test & F1-score  \\
       \Xhline{4\arrayrulewidth}
       \multirow{12}{*}{MNIST 01}& \multirow{2}{*}{4F, 4L} & With $\RX$& 0.9898& 0.9948& 0.9940& 0.9942& 0.9948&0.971& 0.9722 \\ \cline{3-10}
      & & With $\RY$& 0.9918 & 0.9942 & 0.9908 & 0.9905 & 0.9910 & 0.987 & 0.9879 \\ \Xcline{2-10}{3\arrayrulewidth}
      & \multirow{2}{*}{\textbf{4F, 2L}} & With $\RX$& 0.9253& 0.9170& 0.8340& 0.8370& 0.8330 & 0.7855& 0.8243 \\ \cline{3-10}
       && \textbf{With} $\mathbf{\RY}$& 0.9683 & 0.9752 & 0.9738 & 0.9735 & 0.9738 & 0.965 & \textbf{0.9683} \\ \Xcline{2-10}{3\arrayrulewidth}
      & \multirow{2}{*}{6F, 4L} & With $\RX$&0.9610& 0.9798& 0.9808& 0.9808& 0.9798& 0.987& 0.9879 \\ \cline{3-10}
       && With $\RY$& 0.9683 & 0.9798 & 0.9812 & 0.9805 & 0.9792 & 0.982 & 0.9834 \\ \Xcline{2-10}{3\arrayrulewidth}
       & \multirow{2}{*}{\textbf{6F, 2L}} & With $\RX$& 0.6665& 0.6470& 0.6475& 0.6412& 0.6465& 0.6165& 0.7365 \\ \cline{3-10} 
      & & \textbf{With} $\mathbf{\RY}$ & 0.9577 & 0.9688 & 0.9665 & 0.9645 & 0.9663 & 0.9515 & \textbf{0.9567} \\ \Xcline{2-10}{3\arrayrulewidth}
       & \multirow{2}{*}{8F, 4L} & With $\RX$& 0.9670& 0.9708& 0.9722& 0.9720& 0.9725& 0.979& 0.9807 \\ \cline{3-10}
       & & With $\RY$& 0.9775 & 0.9842 & 0.9878 & 0.9860 & 0.9835 & 0.9775 & 0.9794 \\ \Xcline{2-10}{3\arrayrulewidth}
      & \multirow{2}{*}{\textbf{8F, 2L}} & With $\RX$& 0.8147& 0.8240& 0.8323& 0.8300& 0.8363& 0.802 & 0.8440 \\ \cline{3-10}
       & & \textbf{With} $\mathbf{\RY}$& 0.9752 & 0.9765 & 0.9765 & 0.9762 & 0.9778 & 0.9565 & \textbf{0.9610} \\
       \Xhline{4\arrayrulewidth}
        \multirow{12}{*}{MNIST 08} & \multirow{2}{*}{4F, 4L} & With $\RX$& 0.9525& 0.9530& 0.9535& 0.9523& 0.9530 & 0.9508 & 0.9490 \\ \cline{3-10}
      & & With $\RY$& 0.9493 & 0.9513 & 0.9520 & 0.9520 & 0.9515 & 0.9309 & 0.9266 \\ \Xcline{2-10}{3\arrayrulewidth}
      & \multirow{2}{*}{4F, 2L} & With $\RX$& 0.8808 & 0.8942 & 0.8912 & 0.8955 & 0.8960 & 0.9022 & 0.8964 \\ \cline{3-10}
     &  & With $\RY$& 0.9015 & 0.9087 & 0.9090 & 0.9100 & 0.9090 & 0.8751 & 0.8576 \\\Xcline{2-10}{3\arrayrulewidth}
      & \multirow{2}{*}{6F, 4L} & With $\RX$& 0.9320& 0.9395& 0.9443& 0.9437& 0.9453& 0.9498& 0.9496 \\ \cline{3-10}
      & & With $\RY$& 0.9285 & 0.9427 & 0.9415 & 0.9413 & 0.9400 & 0.9334 & 0.9289 \\ \Xcline{2-10}{3\arrayrulewidth}
     & \multirow{2}{*}{6F, 2L} & With $\RX$& 0.9123& 0.9350& 0.9363& 0.9363& 0.9355& 0.9401& 0.9399 \\ \cline{3-10}
     &  & With $\RY$& 0.8972 & 0.9205 & 0.9260 & 0.9273 & 0.9260 & 0.9298 & 0.9250 \\ \Xcline{2-10}{3\arrayrulewidth}
      & \multirow{2}{*}{8F, 4L} & With $\RX$& 0.9390& 0.9323& 0.9303& 0.9320& 0.9325 & 0.9406& 0.9415 \\ \cline{3-10}
      & & With $\RY$& 0.9327 & 0.9427 & 0.9400 & 0.9420 & 0.9403 & 0.9227 & 0.9167 \\ \Xcline{2-10}{3\arrayrulewidth}
     &  \multirow{2}{*}{\textbf{8F, 2}L} & \textbf{With} $\mathbf{\RX}$& 0.8675& 0.9327& 0.9320& 0.9325& 0.9313& 0.9385& \textbf{0.9385} \\ \cline{3-10}
      & & With $\RY$& 0.9095 & 0.9167 & 0.9187 & 0.9223 & 0.9237 & 0.9124 & 0.9047 \\
       \Xhline{4\arrayrulewidth}
    \end{tabular}}
    \normalsize
    \caption{\textit{Benchmarking Simple Angle encoding with $\RX$ and $\RY$ on \textbf{MNIST 01} and \textbf{MNIST08 datasets.}}}
    \label{tab:MNIST01_MNIST08_rx_vs_ry}
\end{table}

\newpage
\subsection*{Graphical results}

\subsubsection*{Malware dataset}
\begin{figure}[H]
    \centering
    \includegraphics[width=0.92\linewidth]{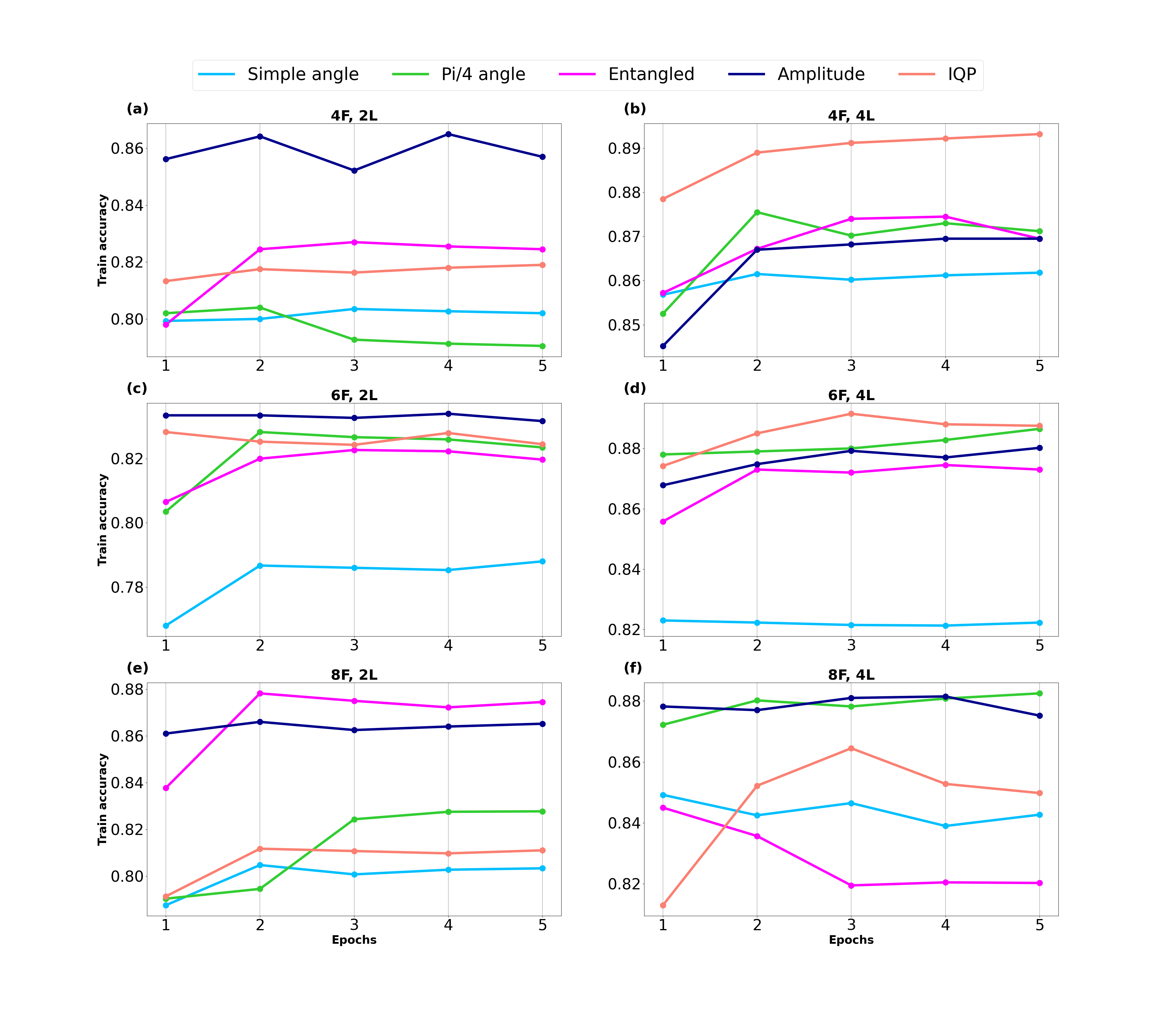}
    \caption{\textit{The five encodings on \textbf{Malware dataset}: training accuracies after 5 epochs, with $N=\{4,6, 8\}, \  M \in \big\{2, 4\big\}$}.}
    \label{fig:Malware_benchmarking the five encodings}
\end{figure}

\subsubsection*{WDBC dataset}
\begin{figure}[H]
    \centering
    \includegraphics[width=0.92\linewidth]{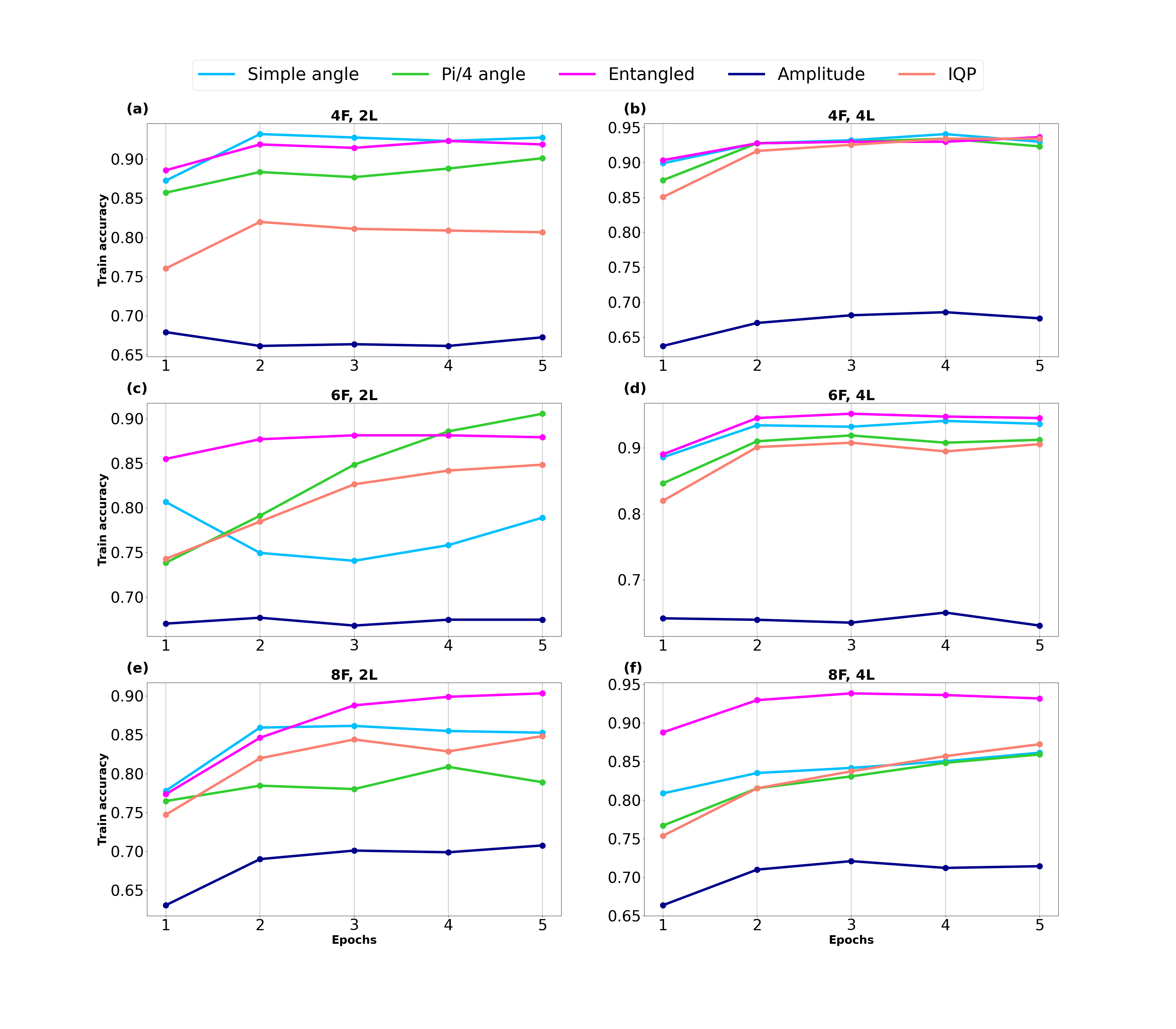}
    \caption{\textit{The five encodings on \textbf{WDBC dataset}: training accuracies after 5 epochs, with $N=\{4,6, 8\}, \  M \in \big\{2, 4\big\}$.}}
    \label{fig:WDBC}
\end{figure}

\subsubsection*{MNIST01 dataset}

\begin{figure}[H]
    \centering
    \includegraphics[width=0.92\linewidth]{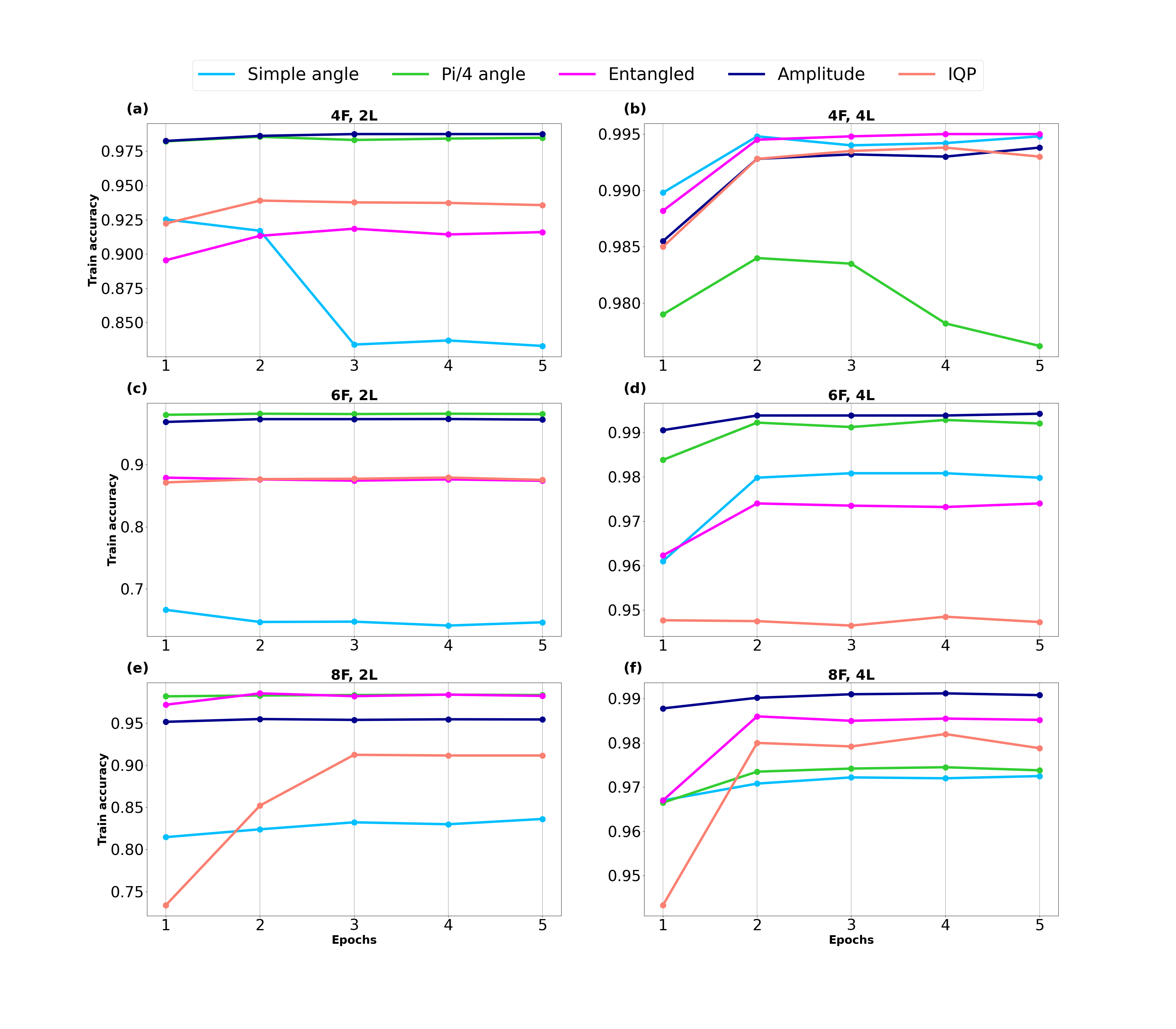}
    \caption{\textit{The five encodings on \textbf{MNIST01 dataset}: training accuracies after 5 epochs, with $N=\{4,6, 8\}, \  M \in \big\{2, 4\big\}$}.}
    \label{fig:MNIST01}
\end{figure}

\subsubsection*{MNIST08 dataset}

\begin{figure}[H]
    \centering
    \includegraphics[width=0.92\linewidth]{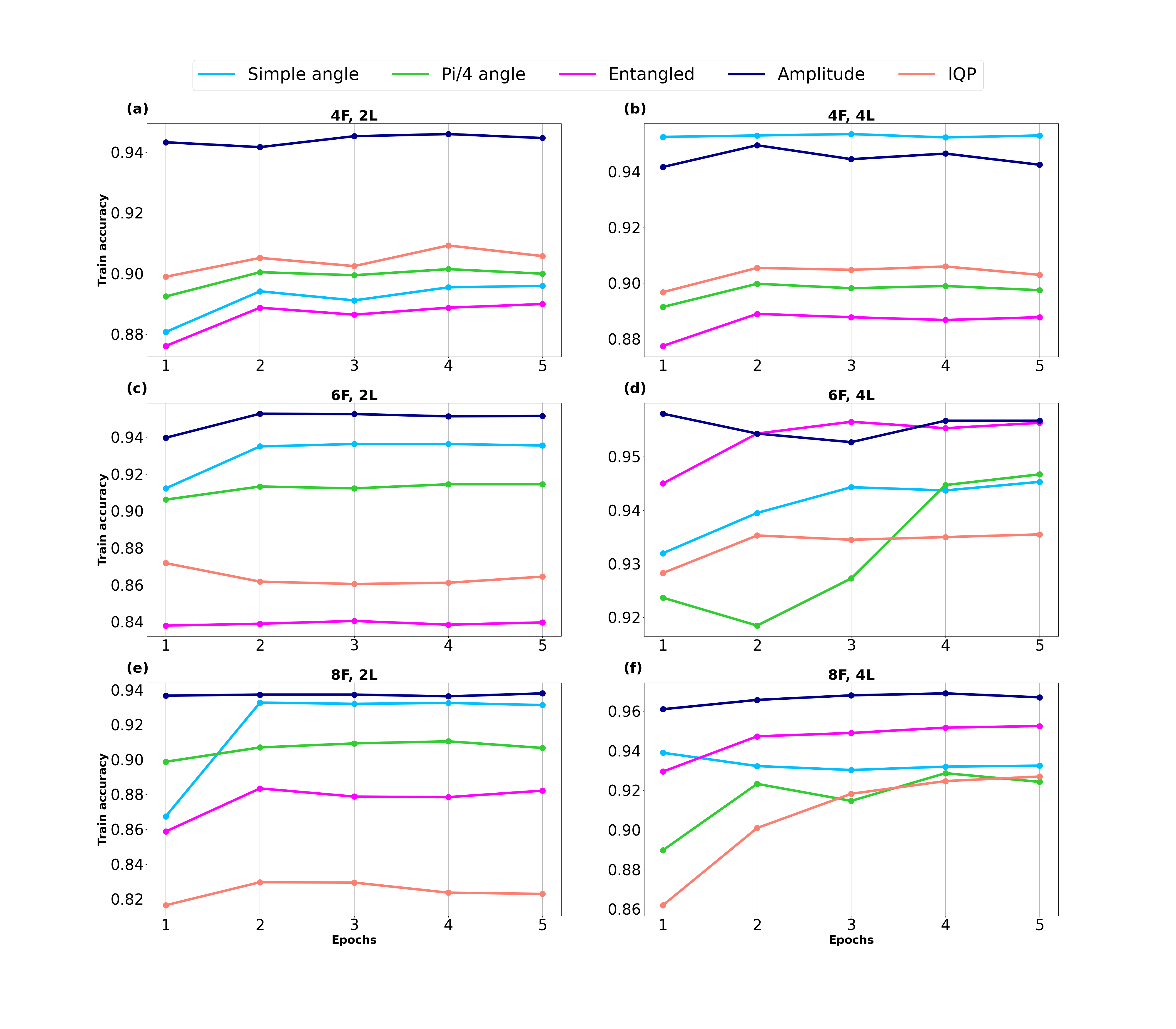}
    \caption{\textit{The five encodings on \textbf{MNIST08 dataset}: training accuracies after 5 epochs, with $N=\{4,6, 8\}, \  M \in \big\{2, 4\big\}$}.}
    \label{fig:MNIST08}
\end{figure}

\end{document}